\documentclass[aps,prl,twocolumn,superscriptaddress,nobibnotes,longbibliography,10pt]{revtex4-2}
\usepackage[utf8]{inputenc}
\usepackage{amsmath,bm}
\usepackage{amsfonts}
\usepackage{amssymb}
\usepackage{xcolor}
\usepackage{braket}
\usepackage{graphicx}
\usepackage{subfigure}
\usepackage{todonotes}
\usepackage{subfigure}
\usepackage{comment}
\usepackage[colorlinks=true, allcolors=blue]{hyperref}

\begin{document}

\title{Encoding classical data into the squeezing of noisy-states for plasmonic communication}

\author{Mehmet Emre Tasgin}
\email[]{Corresponding author: metasgin@hacettepe.edu.tr}
\affiliation{Institute of Nuclear Sciences, Hacettepe University, 06800 Ankara, Turkey}
\date{\today}

\begin{abstract}
Surface plasmon polaritons (SPPs) are known to preserve quantum optical properties—such as squeezing—over distances far exceeding those of classical field amplitudes. However, the surviving squeezing typically becomes so weak that its detection requires prohibitively large numbers of measurements.
 Here we introduce a fundamentally new paradigm for plasmonic communication in which nonclassicality itself carries the information. We (i) encode classical data (bits or dits) directly into the {\it degree of nonclassicality} (e.g., squeezing) of SPPs, thereby enabling information transfer over distances where classical amplitude encoding fails. We further (ii) show that this information can be retrieved from long-lived correlations generated at the readout stage via a beam splitter. Crucially, we demonstrate that (iii) encoding on initially noisy states leads to a counterintuitive enhancement: the encoded information remains accessible after long propagation distances using only a few measurements, outperforming both squeezed vacuum and amplitude-based schemes by orders of magnitude. Finally, (iv) in the THz regime—relevant for graphene and carbon-nanotube platforms at room temperature—we \textit{exploit}, rather than suppress, the intrinsic thermal background, enabling robust, high-bandwidth nanoscale communication.
%
%
%In this work, we show that encoding classical data (bits or dits) into the {\it degree of squeezing} of a single-mode squeezed ``{\it noisy}''  state enables efficient transmission through plasmonic nanowires. Remarkably, this approach allows the encoded information to be recovered even at long propagation distances using only a few measurements—far outperforming both squeezed non-noisy states and amplitude-based encoding schemes. {\red In the THz regime, where graphene and carbon-nanotube SPP platforms operate with significant thermal occupation at room temperature, the protocol exploits the existing noisy background rather than suppressing it, enabling robust high-bandwidth nanoscale interconnects.}
\end{abstract}

\maketitle

Metal nanowires~(MNWs) guide electromagnetic energy in the form of surface plasmon polaritons (SPPs) --coherent charge-density oscillations confined to metal-air or metal-dielectric interfaces~\cite{maier2007plasmonics}. These excitations travel at near-light speeds and operate mainly in the optical and infrared ranges, offering data transfer rates comparable to fiber optics. As a result, MNWs are promising candidates for high-bandwidth nanoscale interconnects~\cite{ozbay2006plasmonics}. However, SPPs suffer from inherent propagation losses that restrict their travel distance to just a few microns~\cite{wiecha2019direct}, rendering them unsuitable even for intermediate interconnects within a single CPU core~\footnote{Embedding MNWs in dielectric environments can extend the range to millimeters-known as long-range SPPs~\cite{berini2009long}. However, these structures are bulkier and reduce integration density.}.
In addition, the limited development of enabling technologies --such as coherent SPP sources (e.g., spasers~\cite{noginov2009demonstration}) and efficient plasmonic detectors --remains a challenge. Among all limitations, however, the most critical one is the short SPP propagation length.

%
%
%Metal nanowires (MNWs) guide electromagnetic energy in the form of surface plasmon polaritons (SPPs)—coherent charge-density oscillations confined to the metal/air or metal/dielectric interfaces~\cite{maier2007plasmonics}. These excitations travel at near-light velocities and operate predominantly in the optical and infrared spectra, offering potential for high-bandwidth data transfer comparable to fiber optics. This makes MNWs attractive candidates for nanoscale interconnects with large bandwidth capacities~\cite{ozbay2006plasmonics}. However, SPPs suffer from intrinsic propagation losses that limits the SPP propagation length to micron~\cite{wiecha2019direct}. This by itself limits the implementation to interconnects that transfer data among different components of the CPU~\footnote{When MNW is embedded in a dielectric environment, the range can extend to millimeters ---long-range SPPs~\cite{berini2009long}. However, they occupy much larger space compared to bare MNWs limiting the integribility.}. Moreover, limited maturity of supporting technologies such as coherent SPP sources (e.g., spasers~\cite{noginov2009demonstration}) and efficient plasmonic detectors remain as other significant challenges. Yet, the most important limitation origins from the short propagation length.

Interestingly, despite rapid intensity attenuation, the quantum optical properties of SPPs ---governed by quantum noise $\delta\hat{a} = (\hat{a} - \langle \hat{a} \rangle)$--- can persist over much longer distances. Experiments involving MNWs and metallic nanostripes have shown that losses in quadrature squeezing are markedly smaller than those in SPP intensity~\cite{huck2009demonstration}. Furthermore, photon number correlations, quantified by the second-order coherence function $g^{(2)}$, have been observed to remain constant along metallic nanostructures when heralded single-photon sources are used~\cite{di2012quantum,hong2024nonclassical}. This apparent robustness arises because thermal noise at optical frequencies is effectively negligible, allowing quantum features to survive even in lossy environments~\cite{huck2009demonstration,di2012quantum,fasel2005energy,fasel2006quantum}.

The degradation of quantum features along MNWs can be modeled using a beam splitter-like transformation~\cite{huck2009demonstration}
\begin{equation}
	\hat{a}_{\rm \scriptscriptstyle T} = \eta \: \hat{a} + \sqrt{1-\eta} \: \hat{b}_{\rm vac},
	\label{BS}
\end{equation}
where $\eta = e^{-L/L_0}$ is the transmission coefficient of the SPP intensity over a distance $L$, and $L_0$ is the characteristic propagation length—typically microns in the optical regime and millimeters in the infrared (long-range SPPs)~\cite{berini2009long}. Ref.~\cite{huck2009demonstration} demonstrates that Eq.~(\ref{BS}) successfully governs the relatively lower degrading in the squeezing~(24\%) despite the 67\% loss in the intensity. It also explains the constant behavior of $g^{(2)}$~\cite{di2012quantum,hong2024nonclassical,tasgin2026}.
The observed constancy of $g^{(2)}(0)$ stems from its normalized dependence on intensity, unlike squeezing, which scales linearly with amplitude. However, it is important to note that demonstration of  the constancy of $g^{(2)}(0)$ crucially depends on the presence of heralded photon sources~\cite{di2012quantum}.

While squeezing can persist over longer distances than the SPP amplitude, reliably detecting small residual squeezing values --such as a squeezing of only 0.999 after propagation over a distance of several $L=10L_0$-- requires averaging over millions to billions of state copies. Such a demand for large statistical ensembles limits practical utility. However, if squeezing levels could be distinguished with just a few measurements, one could exploit the wide bandwidth of SPPs for efficient classical data transfer using quantum features.

\begin{figure}%[t]
	\centering
	\includegraphics[width=1\linewidth, trim={3.8cm 1cm 0 1cm},clip]{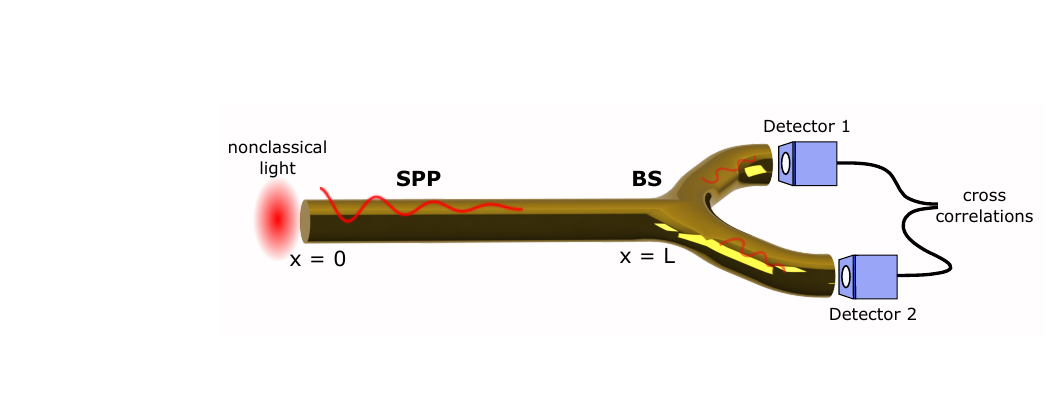}
	% Answer: [trim={left bottom right top},clip]
	\caption{{\it Encoding classical data into squeezing degrees of SPPs.}
		We encode classical bits into the degree of squeezing of surface plasmon polaritons (SPPs): $r_0 = 0$ (representing 0-bit) and $r_1 = 0.574$ (1-bit). Instead of conventional squeezed vacuum states, we use single-mode squeezed noisy states (SMSNSs), which are more robust against propagation losses~\cite{PS_THz_advantange,PS_numner_of_photons}. The SPP, prepared in this squeezed noisy state, travels from position $x=0$ to $x=L$ along the nanowire. At $x=L$, it encounters a plasmonic beam splitter (BS), which splits the SPP into two modes and introduces correlations between them. The encoded squeezing information is retrieved by measuring the quadrature cross-correlations at the output ports—similar to techniques used in quantum illumination studies~\cite{gundogan2025enhanced,huang2023photonic,tan2008quantum,zhuang2021quantum,guha2009gaussian,zhuang2017optimum,shi2023fulfilling,reichert2023quantum,jeon2025single,karsa2024quantum,barzanjeh2015microwave,qian2023quantum}.}
	\label{fig1}
\end{figure}

In this paper, we propose transferring classical data by {\it encoding it into quantum noise features}, exploiting the fact that nonclassical properties survive significantly longer than intensity profiles—and, in the case of $g^{(2)}(0)$, may not decay at all~\cite{di2012quantum,tasgin2026,hong2024nonclassical}. Rather than directly measuring squeezing after propagation~\cite{di2012quantum,huck2009demonstration,fasel2006quantum}, we employ a plasmonic beam splitter~\cite{PS_squeezing_detection} to access the even longer-lived correlations at the \textit{readout port}. In contrast to quantum illumination (QI) schemes~\cite{gundogan2025enhanced,huang2023photonic,tan2008quantum,zhuang2021quantum,barzanjeh2015microwave}, which rely on source-generated two-mode squeezing, we generate a single-mode squeezed SPP and infer its nonclassicality at the \textit{readout port} via correlations, without any return signal (see Fig.~\ref{fig1}).

%
%{\red In this paper, we propose to transfer classical data {\it by encoding it into quantum noise features}, as nonclassical properties are known to survive significantly longer (and in the case of $g^{(2)}(0)$, not decay at all~\cite{di2012quantum,tasgin2026,hong2024nonclassical}) compared to intensity profiles~\cite{di2012quantum,huck2009demonstration,fasel2006quantum}. Instead of directly reading the squeezing at a distance~\cite{di2012quantum,huck2009demonstration,fasel2006quantum}, we use a plasmonic beam splitter~\cite{PS_squeezing_detection} and measure the even longer surviving correlations at the \textit{readout port}. In difference to QI techniques~\cite{gundogan2025enhanced,huang2023photonic,tan2008quantum,zhuang2021quantum,barzanjeh2015microwave}, which uses source-generated two-mode squeezing, we generate a single-mode squeezed SPP and infer its nonclassicality at the \textit{readout port} from correlations, with no return signal, see Fig.~\ref{fig1}.}

We propose to encode classical data into discrete {\it squeezing} values—for example, $r_0=0$ for a 0-bit, $r_1=0.15$ for a 1-bit, $r_2=0.30$ for a 2-dit, and so on. Such encoding can be performed electrically and within very short time sequences~\footnote{In principle in picosecond sequences, limited by the control solid-state device.} in integrated circuits using Fano resonances~\cite{gunay2023demand} and the Stark effect. Moreover, since squeezing-based encoding can in principle be implemented independently across different spectral bands, there is no fundamental obstacle to multi-wavelength quantum data transfer using this method. In Ref.~\cite{tasginSqueezingMultiplexed}, we propose an electrically tunable, multi-frequency squeezing-based platform for such applications.

Although we focus here on squeezing-degree encoding—owing to the analytical convenience of Gaussian states—encoding classical data into $g^{(2)}(0)$ values holds the  {\it potential for a transformative shift}, as experiments and simple analytical treatments show that $g^{(2)}(0)$ remains invariant during SPP  propagation~\cite{PS_g2_encoding}. Moreover, while we consider optical SPPs, this approach becomes even more compelling in the THz regime, for example in graphene or carbon nanotube (CNT) platforms~\cite{ukhtary2020surface}, where \textit{free} thermal photons naturally exist at room temperature~\cite{riedinger2018remote,abdi2015entangling,balram2017acousto,zhao2023electro,huang2023photonic}~\cite{PS_room_T_operation}. In such systems, the background noise can be harnessed as a resource, potentially enabling data transfer rates exceeding those of conventional electronic approaches~\cite{PS_THz_advantange}.

%{\red Although we focus on squeezing-degree encoding here—because Gaussian states are convenient to analyze—significant potential exists in encoding classical data into $g^{(2)}(0)$ values, since experiments and simple analytical treatments show that $g^{(2)}(0)$ does not change during SPP propagation~\cite{PS_g2_encoding}. In addition, although we consider optical SPPs in this work, as discussed in detail at the end of the text, such encoding becomes even more promising for SPPs operating in the THz regime, for example on graphene or carbon nanotube~(CNT) platforms~\cite{ukhtary2020surface}, where \textit{free} thermal photons naturally exist at room temperature~\cite{riedinger2018remote,abdi2015entangling,balram2017acousto,zhao2023electro,huang2023photonic}~\cite{PS_room_T_operation}. In such systems, our method can use the background noise as a resource and potentially enable higher data transfer rates than electronic approaches~\cite{PS_THz_advantange}.}

 Here, we present a simple and effective method for detecting squeezing-encoded data with only a few measurements, even after substantial propagation loss.  The key idea is to encode information into nonclassicality and to deliberately prepare the single-mode (SM) system in a highly noisy Gaussian state (with mean photon number $\bar{n}_{\rm p}=10{,}000$) \textit{prior} to applying single-mode squeezing (SMS). Specifically, the squeezing operation is performed on this \textit{intentionally} noisy state, which is then transmitted through a channel that introduces additional noise.

%{\red Here,} we present a simple and effective technique that enables detection of squeezing-encoded data with only a few measurements—even after substantial propagation loss. {\red The key ideas are to encode data into nonclassicality and to deliberately prepare the single-mode~(SM) system in a highly noisy {\red Gaussian} state (with mean photon number $\bar{n}_{\rm p}=10{,}000$) before applying single-mode squeezing~(SMS). {\red That is, we perform the single-mode squeezing operation on an \textit{intentionally} prepared noisy state and then transmit it through a channel that introduces its own noise.}

 While squeezed vacuum and squeezed-noisy SPP states exhibit similar detection rates \textit{before} entering the channel, the squeezed-noisy state is significantly more robust to propagation noise in Eq.~(\ref{BS}). This robustness arises from its large photon-number background and correspondingly {\it enhanced off-diagonal} covariance elements. When squeezing is applied to an initially noisy Gaussian state, the preparation noise {\it co-amplifies} with the inter-mode covariance, leaving the normalized correlation signal largely unaffected as the noise increases. After transmission, this enhanced covariance becomes more resilient to the added (constant) channel noise, enabling the squeezed-noisy state to maintain a signal-to-noise ratio (SNR) orders of magnitude higher than that of the squeezed vacuum state~\footnote{See Ref.~\cite{gundogan2025enhanced,huang2023photonic} for details of the technique studied in the context of quantum imaging and work extraction.} \cite{PS_numner_of_photons}.

Such a noisy optical state can be generated using amplitude or phase modulation devices, or parametric amplifiers~\cite{ScullyZubairyBook,gerry2023introductory,guo2023chaos,chamoun2017aircraft,zao2009design,chamoun2016pseudo,capmany2011quantum}. Moreover, it naturally exists at room temperature in devices~\cite{ukhtary2020surface,huang2023photonic} operating, for example, with THz-regime SPPs~\cite{PS_numner_of_photons}.

The read-out scheme is as follows. An SPP is excited by nonclassical light~\cite{huck2009demonstration,di2012quantum}, into which classical data is encoded in the squeezing degree~\footnote{Once spaser technology reaches sufficient maturity, direct generation of nonclassical SPPs may also become possible~\cite{noginov2009demonstration,gunay2023demand}.}, as illustrated in Fig.~1. The SPP then propagates along a metallic nanowire (MNW) from $x=0$ to $x=L=10L_0$.
At the readout port, the SPP is split by a plasmonic 50:50 beam splitter (BS)~\cite{han2010wideband,heeres2013quantum,PS_squeezing_detection}. Although the squeezing at 
$x=L$ is largely obscured by fluctuations, it gives rise to measurable two-mode correlations (without entanglement) at the BS outputs~\cite{gundogan2025enhanced,tan2008quantum,zhuang2021quantum,guha2009gaussian,zhuang2017optimum,shi2023fulfilling,reichert2023quantum,jeon2025single,karsa2024quantum,barzanjeh2015microwave,qian2023quantum}~\cite{PS_squeezing_detection}.

The encoded squeezing, embedded in the noisy background ($\bar{n}_{\rm p}=10,000$), gives rise to observable quadrature correlations at the BS outputs~\cite{kim2002entanglement,tan2008quantum,torrome2024advances}. This is because a BS transforms single-mode squeezing into entanglement or, more generally, measurable correlations between its outputs~\cite{ge2015conservation,braunstein2005quantum,ou1987detection}. The strength of these correlations grows with the degree of squeezing~\cite{asboth2005computable,ou1987detection}. While thermal noise typically degrades entanglement in two-mode squeezed states~\cite{duan2000inseparability,tasgin2019anatomy}, quadrature correlations can survive and remain detectable --similar to what is exploited in quantum illumination protocols~\cite{gundogan2025enhanced,tan2008quantum,zhuang2021quantum,barzanjeh2015microwave}. As a result, the signal-to-noise ratio~(SNR) of these correlations can be sufficiently high that even at propagation lengths as large as $L=10L_0$ (corresponding to transmittivity $\eta=e^{-10}=0.45\times 10^{-4}$), only two copies ($M=2$) of the single-mode squeezed noisy state ($\bar{n}_{\rm p}=10,000$ and squeezing $r=0.576$~\cite{PS_squeezingNp}) is enough for detection. This is in stark contrast to the case of a squeezed vacuum state ($\bar{n}_{\rm p}=0$), where $M=10^5$ measurements are required to extract the same squeezing information at comparable distances.~\footnote{
There are three conceptually distinct methods for extracting squeezing information after SPP propagation:  
(i) Directly measuring noise in the single-mode SPP state after propagation over $L = 10L_0$;  
(ii) Splitting the same single-mode state using a beam splitter and then measuring output correlations ---this approach significantly reduces the number of copies needed;  
(iii) Using a squeezed noisy state (rather than a squeezed vacuum) and performing correlation measurements after beam splitting. This last method reduces the number of required measurements by a factor of $10^{-5}$ to $10^{-4}$ compared to method (ii)~\cite{PS_numner_of_photons}.
}

\begin{figure}%[t]
	\centering
	\includegraphics[width=1\linewidth]{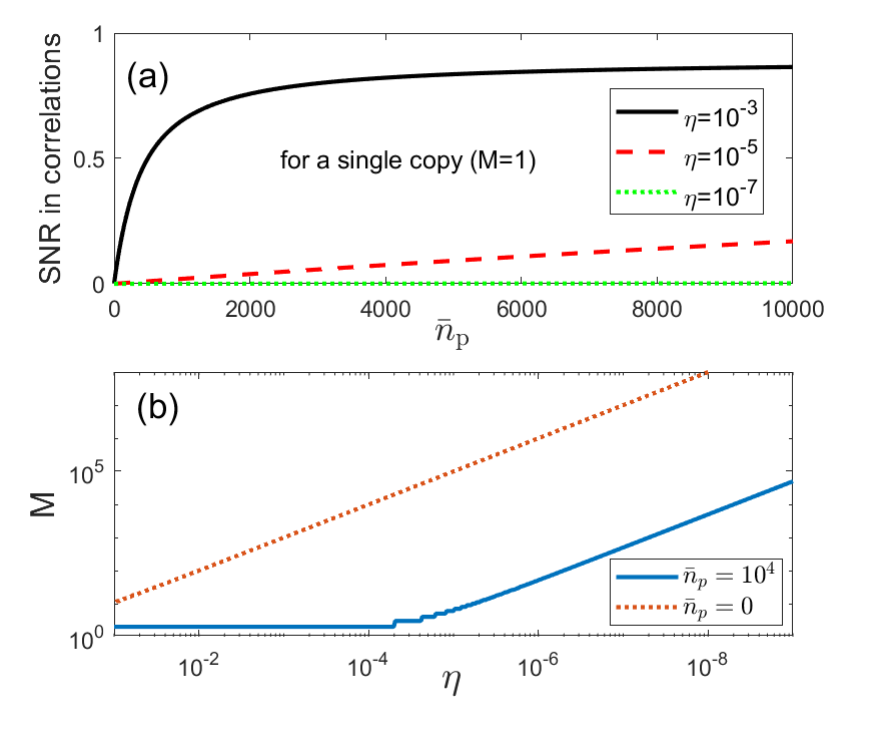}
	\caption{{\it (a)  Signal-to-noise ratio (SNR) vs. preparation noise.}
		Here, we analyze the SNR of correlation measurements for a single-copy ($M = 1$) of an SMSNS with squeezing $r = 0.576$~\cite{PS_squeezingNp}. We find that increasing the initial preparation noise $\bar{n}_{\rm p}$ enhances the detectability of the correlation signal, making it easier for detectors to resolve the squeezing.
		{\it (b) Copy count needed for detection.}
		This panel shows the minimum number of copies required to distinguish between the absence ($r_0=0$) and presence ($r_1=0.576$) of squeezing. Remarkably, even after long-distance propagation ($L = 10L_0$), where the SPP transmission drops to $\eta = e^{-10} = 0.45 \times 10^{-4}$, the presence of squeezing can be identified using as few as two copies.}
	\label{fig2}
\end{figure}

{\it Mechanism}.--- To clarify the mechanism behind our approach, we first isolate the core idea by removing the propagation stage from the setup in Fig.~1. That is, we conceptually place the beam splitter~(BS) at the origin, $x=0$, before any propagation occurs. The key observation is that the signal-to-noise ratio~(SNR) of the quadrature correlations at the BS output remains essentially the same~\cite{gundogan2025enhanced}, whether the initial state {---before the single-mode squeezing operation is applied---} is vacuum ($\bar{n}_{\rm p}=0$) or highly noisy ($\bar{n}_{\rm p}=10{,}000$). However, this equivalence becomes a clear advantage once propagation between $x=0$ and $x=L$ is included. In that case, preparing the state with noise enhances the visibility of the correlations after loss by orders of magnitude.

%{\it Mechanism}.--- To clarify the mechanism behind our approach, let us first isolate the core idea by removing the propagation stage from the setup in Fig.~1. That is, we conceptually place the beam splitter~(BS) at the origin, $x=0$, before any propagation takes place. The key observation is that the signal-to-noise ratio~(SNR) in the quadrature correlations at the BS output remains effectively unchanged whether the initial state {\red {---before the single-mode squeezing operation is applied---}} is vacuum ($\bar{n}_{\rm p}=0$) or highly-noisy ($\bar{n}_{\rm p}=10,000$). However, this equivalence becomes a major advantage when we reintroduce propagation between $x=0$ and $x=L$. In that case, the noisy preparation significantly enhances the visibility of correlations after loss.

Including the propagation stage, the noise evolution of the single-mode state is captured by the covariance matrix~(CM) transformation
\begin{equation}
 V(L) = \eta \: V_{\rm sqz} + (1-\eta)\:V_{\rm vac},
\end{equation}
where $V_{\rm sqz}$ is the CM of the initial squeezed state and $V_{\rm vac}$ corresponds to the vacuum noise injected due to propagation loss~\cite{huck2009demonstration}. Crucially, both the squeezed-vacuum state ($\bar{n}_{\rm p}=0$) and the squeezed noisy state ($\bar{n}_{\rm p}=10,000$) experience the same absolute amount of vacuum noise injection ---equivalent to a $1/2$-unit of added noise per quadrature. However, in the noisy case, this added vacuum noise is relatively small compared to the already large initial Gaussian noise and the off-diagonal elements of $V_{\rm sqz}$. In contrast, when starting from a vacuum state, the same vacuum noise becomes dominant, effectively drowning out any signature of squeezing~\cite{PS_THz_advantange}. As a result, after propagation over a distance $L=10L_0$~(corresponding to $\eta=e^{-10}=0.45\times 10^{-4}$), the squeezed noisy state still exhibits clear quadrature correlations at the BS output ---detectable even with a few copies. This behavior is illustrated in Fig.~2a.

 Figure 2a shows that the SNR of the measured correlations improves significantly as the preparation noise $\bar{n}_{\rm p}$ increases. To detect low-SNR signals, one can apply central limit theorem arguments and use hypothesis testing techniques, such as those employed in quantum illumination~\cite{barzanjeh2015microwave,barzanjeh2020microwave}. For example, detecting a correlation signal with an SNR of 0.2 requires approximately $M=1/{\rm SNR}=5$ state copies. In Fig.~2b, we show the number of copies $M$ required to distinguish between the absence~($r_0=0$) and presence ($r_1=0.576$, equivalent to 10 dB~\cite{PS_squeezingNp}) of squeezing. The results reveal that the noisy squeezed state ($\bar{n}_{\rm p}=10,000$) drastically outperforms the squeezed-vacuum state~($\bar{n}_{\rm p}=0$), especially at longer propagation lengths. For example, with $\eta=e^{-10}=0.45\times 10^{-4}$), only two copies of the squeezed noisy state suffices to resolve the presence of squeezing, whereas the squeezed-vacuum case requires 5 orders of magnitude more copies. This performance gap grows with increasing distance, due to the exponential dependence of $\eta$ on propagation length.

\begin{figure}%[t]
	\centering
	\includegraphics[width=1\linewidth]{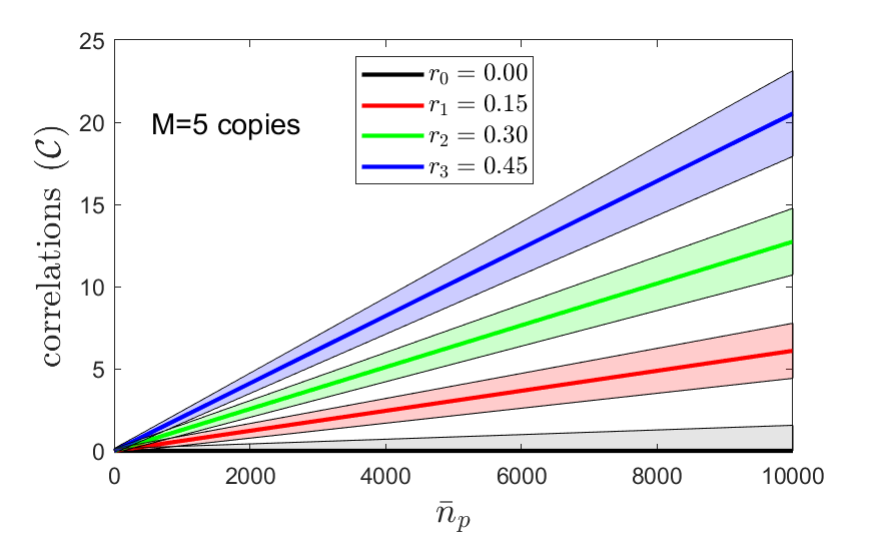}
	\caption{{\it Multi-level data encoding with squeezing.}
		We extend the scheme to a larger alphabet where multiple classical symbols are encoded using different squeezing levels (e.g., 0-dit, 1-dit, 2-dit, etc.). The shaded regions in the figure represent the correlation signal and noise associated with each squeezing-encoded symbol. This demonstrates the feasibility of high-density classical data encoding into squeezing degrees, paving the way for wavelength-multiplexed and noise-resilient plasmonic communication.}
	\label{fig3}
\end{figure}

In our simulations, the correlation observable is
\begin{equation}
{\cal C}=\langle \hat{a}_1 \hat{a}_2 +  \hat{a}_2^\dagger \hat{a}_1^\dagger \rangle  =\hat{x}_1 \hat{x}_2 - \hat{p}_1 \hat{p}_2,
\end{equation}
which captures the entanglement-like correlations created at the BS by a single-mode squeezed input~\cite{asboth2005computable,torrome2024advances}. We focus on zero-mean states that possess experimental viability. Importantly, this observable is sensitive to the nonclassical structure in number-non-conserving terms that are characteristic of squeezing-converted correlations. Such correlations are measured using optical parametric amplifier, phase-conjugate, sum-frequency generating, correlation-to-displacement receivers or hetero-homodyne receiver with sequential detection methods~\cite{guha2009gaussian,zhuang2017optimum,shi2023fulfilling,reichert2023quantum,karsa2024quantum,jeon2025single}.

Finally, we explore the feasibility of using a larger encoding alphabet. Specifically, we consider squeezing values corresponding to multiple logical states  $r_0=0$~(0-dit), $r_1=0.1$~(1-dit),$r_2=0.2$~(2-dit) and $r_3=0.3$~(3-dit). In Fig.~3, we plot both the correlations ${\cal {C}}$ and their corresponding SNRs for each of these encoding levels at a propagation length of $L=10 L_0$. The results confirm that a multi-level~(multi-dit) squeezing-based data encoding scheme is also viable, and that using squeezed-noisy states  significantly enhances distinguishability across all symbols in the alphabet.

%%%%%%%%%%%%%%%%%%%%%%%%%%%%%%%%%%%%%%%%%%%%%%%%%%%%%%%%%%%%%%%%%%%%%%%%%%%%%%%%%%%%%%%%%%%%%%%%%%%%%%%%%%%%%%%%%%%%%%%%%%%%%%%%%

{\it  Summary, Conclusions and Outlook}.--- We introduce a method to encode classical information into surface plasmon polaritons~(SPPs) using the degree of squeezing, and more generally nonclassicality, as a data carrier. Different squeezing values represent classical symbols~(dits), forming a squeezing-based communication alphabet. Although squeezing is known to survive longer than amplitude in plasmonic nanowires~(NWs), it still decays significantly. Even for a moderate propagation length of $L=10L_0$, recovering small residual squeezing typically requires millions to billions of repeated measurements.

 Unlike Refs.~\cite{huck2009demonstration,di2012quantum,fasel2005energy,fasel2006quantum}, we propose to use a plasmonic beam splitter and detect the correlations~\cite{PS_squeezing_detection} at its outputs—employing techniques developed for quantum illumination~\cite{gundogan2025enhanced,huang2023photonic,tan2008quantum,zhuang2021quantum,guha2009gaussian,zhuang2017optimum,shi2023fulfilling,reichert2023quantum,jeon2025single,karsa2024quantum,barzanjeh2015microwave,qian2023quantum}—instead of directly measuring the small remaining squeezing, which is easily washed out by propagation noise. We further encode the squeezing on an \textit{intentionally} noisy initial state, which protects the SNR against the noise introduced during channel propagation~\cite{gundogan2025enhanced,huang2023photonic}.
This approach can even exploit the freely existing thermal photons~\cite{huang2023photonic}, for example in the THz regime at room temperature, to enhance robustness against channel noise~\cite{PS_numner_of_photons}.

Our key result shows that, by performing single-mode squeezing on an initially noisy state and applying beam-splitter-based correlation detection, the required number of measurements can be drastically reduced. In particular, even after long propagation distances (e.g., $L=10L_0$, where the transmission amplitude is $\eta=e^{-10}=0.45\times10^{-4}$), the encoded squeezing can be decoded using only {\it two} copies of the state. This enables reliable detection of classical information without large ensembles, overcoming a major limitation of squeezed-state propagation in plasmonic systems.

Although we focus on optical SPPs, the strongest advantage appears for high-capacity SPP data transfer in platforms such as graphene or carbon nanotubes~(CNTs), which support SPPs in the THz regime~\cite{ukhtary2020surface,huang2023photonic}, where thermal noise is already present at room temperature. In this regime, SPPs are inherently noisy and classical encoding becomes strongly limited. In our approach, single-mode squeezing is applied directly to this already noisy state. A simple comparison of total photon numbers with a classical source may suggest similar energy costs. However, in the THz regime at room temperature, the background noise is naturally present. A classical SPP signal must be generated by extracting work from this noisy environment, whereas in our method the squeezing operation acts directly on (exploits) the existing noisy state (photons). Therefore, our technique overcomes the GHz-range limits of traditional electronics and the limitations of standard encoding schemes by exploiting electromagnetic field oscillations to enable high-bandwidth SPP communication in the noisy THz regime, without requiring additional energy beyond that needed to generate the same single-mode squeezing. This points to a promising direction for THz circuit technologies~\cite{PS_THz_advantange} and emerging 6G bands for open-air THz communication.

In addition, the nonclassicality encoding developed here may have important implications for $g^{(2)}(0)$-based encoding in the optical regime. Experiments and analytical studies show that $g^{(2)}(0)$ nonclassicality remains unchanged during SPP propagation~\cite{di2012quantum,hong2024nonclassical,tasgin2026}, even under strong intensity decay.  While we focus on quadrature-squeezing encoding, $g^{(2)}(0)$-based approaches may offer \textit{transformative} routes for data transfer. The main challenge is accessing $g^{(2)}(0)$ at very low intensities, which may be addressed using beam-splitter correlation measurements~\cite{PS_squeezing_detection} or by converting number-squeezing into quadrature-squeezing detectable at vanishing intensity~\cite{PS_g2_encoding}.

%{\red The nonclassicality-based encoding developed here may also extend to $g^{(2)}(0)$-based schemes in the optical regime. Previous studies show that $g^{(2)}(0)$ nonclassicality is preserved during SPP propagation~\cite{di2012quantum,hong2024nonclassical,tasgin2026}, even under strong intensity decay. While we focus on quadrature-squeezing encoding, $g^{(2)}(0)$-based approaches may offer additional (transformative) routes for data transfer. The main challenge is accessing $g^{(2)}(0)$ at very low intensities. This may be addressed using beam-splitter correlation measurements~\cite{PS_squeezing_detection} or by converting number-squeezing into quadrature-squeezing detectable at vanishing intensity~\cite{PS_g2_encoding}.}

%	
%	Although we focus on quadrature-squeezing encoding, $g^{(2)}(0)$-based encoding may offer further opportunities for data transfer. The main challenge is reading $g^{(2)}(0)$ at very low intensities. One possible solution is to use the same beam-splitter correlation method employed here~\cite{PS_squeezing_detection} to detect number correlations or transforming the number-squeezing into quadrature-squeezing that can be measured with zero intensity~\cite{PS_g2_encoding}.}
%

Although integrated coherent SPP sources and fully plasmonic detection schemes are still technologically immature, rapid progress in plasmonic integration is likely to make such systems feasible in the near future. In this setting, our approach enables wideband classical data transmission over metal nanowires, CNTs, and graphene platforms, with information encoded in the squeezing degree.
Moreover, in Ref.~\cite{tasginSqueezingMultiplexed}, we propose an on-chip, electrically tunable device that can generate different squeezing levels at different wavelengths. Such a platform could function as a squeezing-based wavelength-division multiplexing (WDM) unit, substantially increasing the total data throughput of plasmonic interconnects. 
 
%
%Although integrated, coherent SPP sources and all-plasmonic detection schemes remain technologically immature today, we anticipate that rapid advancements in plasmonic integration will soon make such systems viable. In this context, our approach opens the door to wideband classical data transmission over metal nanowires, CNTs, graphene etc. with information encoded in squeezing amplitudes. Furthermore, in Ref.~\cite{tasginSqueezingMultiplexed}, we propose an on-chip, electrically tunable device capable of generating different squeezing levels at different wavelengths. Such a device could serve as a squeezing-based wavelength-division multiplexing (WDM) unit, significantly enhancing the total data throughput of plasmonic interconnects.

We gratefully thank Rasim Volga Ovali for preparing Fig.~1.

%{\color{red} While we confine ourselves to the correlation enhancement for quadrature squeezing here, obviously, the smaller noise transfer also   }

% 
%\begin{acknowledgments}
%MET gratefully thanks Ozan Ari for illuminating discussions.
%\end{acknowledgments}
%
\bibliography{bibliography}

\begin{thebibliography}{65}%
\makeatletter
\providecommand \@ifxundefined [1]{%
 \@ifx{#1\undefined}
}%
\providecommand \@ifnum [1]{%
 \ifnum #1\expandafter \@firstoftwo
 \else \expandafter \@secondoftwo
 \fi
}%
\providecommand \@ifx [1]{%
 \ifx #1\expandafter \@firstoftwo
 \else \expandafter \@secondoftwo
 \fi
}%
\providecommand \natexlab [1]{#1}%
\providecommand \enquote  [1]{``#1''}%
\providecommand \bibnamefont  [1]{#1}%
\providecommand \bibfnamefont [1]{#1}%
\providecommand \citenamefont [1]{#1}%
\providecommand \href@noop [0]{\@secondoftwo}%
\providecommand \href [0]{\begingroup \@sanitize@url \@href}%
\providecommand \@href[1]{\@@startlink{#1}\@@href}%
\providecommand \@@href[1]{\endgroup#1\@@endlink}%
\providecommand \@sanitize@url [0]{\catcode `\\12\catcode `\$12\catcode
  `\&12\catcode `\#12\catcode `\^12\catcode `\_12\catcode `\%12\relax}%
\providecommand \@@startlink[1]{}%
\providecommand \@@endlink[0]{}%
\providecommand \url  [0]{\begingroup\@sanitize@url \@url }%
\providecommand \@url [1]{\endgroup\@href {#1}{\urlprefix }}%
\providecommand \urlprefix  [0]{URL }%
\providecommand \Eprint [0]{\href }%
\providecommand \doibase [0]{https://doi.org/}%
\providecommand \selectlanguage [0]{\@gobble}%
\providecommand \bibinfo  [0]{\@secondoftwo}%
\providecommand \bibfield  [0]{\@secondoftwo}%
\providecommand \translation [1]{[#1]}%
\providecommand \BibitemOpen [0]{}%
\providecommand \bibitemStop [0]{}%
\providecommand \bibitemNoStop [0]{.\EOS\space}%
\providecommand \EOS [0]{\spacefactor3000\relax}%
\providecommand \BibitemShut  [1]{\csname bibitem#1\endcsname}%
\let\auto@bib@innerbib\@empty
%</preamble>
\bibitem [{\citenamefont {Maier}\ \emph {et~al.}(2007)\citenamefont {Maier}
  \emph {et~al.}}]{maier2007plasmonics}%
  \BibitemOpen
  \bibfield  {author} {\bibinfo {author} {\bibfnamefont {S.~A.}\ \bibnamefont
  {Maier}} \emph {et~al.},\ }\href@noop {} {\emph {\bibinfo {title}
  {Plasmonics: fundamentals and applications}}},\ Vol.~\bibinfo {volume} {1}\
  (\bibinfo  {publisher} {Springer},\ \bibinfo {year} {2007})\BibitemShut
  {NoStop}%
\bibitem [{\citenamefont {Ozbay}(2006)}]{ozbay2006plasmonics}%
  \BibitemOpen
  \bibfield  {author} {\bibinfo {author} {\bibfnamefont {E.}~\bibnamefont
  {Ozbay}},\ }\bibfield  {title} {\bibinfo {title} {Plasmonics: merging
  photonics and electronics at nanoscale dimensions},\ }\href@noop {}
  {\bibfield  {journal} {\bibinfo  {journal} {Science}\ }\textbf {\bibinfo
  {volume} {311}},\ \bibinfo {pages} {189} (\bibinfo {year}
  {2006})}\BibitemShut {NoStop}%
\bibitem [{\citenamefont {Wiecha}\ \emph {et~al.}(2019)\citenamefont {Wiecha},
  \citenamefont {Al-Daffaie}, \citenamefont {Bogdanov}, \citenamefont
  {Thomson}, \citenamefont {Yilmazoglu}, \citenamefont {Kuppers}, \citenamefont
  {Soltani},\ and\ \citenamefont {Roskos}}]{wiecha2019direct}%
  \BibitemOpen
  \bibfield  {author} {\bibinfo {author} {\bibfnamefont {M.~M.}\ \bibnamefont
  {Wiecha}}, \bibinfo {author} {\bibfnamefont {S.}~\bibnamefont {Al-Daffaie}},
  \bibinfo {author} {\bibfnamefont {A.}~\bibnamefont {Bogdanov}}, \bibinfo
  {author} {\bibfnamefont {M.~D.}\ \bibnamefont {Thomson}}, \bibinfo {author}
  {\bibfnamefont {O.}~\bibnamefont {Yilmazoglu}}, \bibinfo {author}
  {\bibfnamefont {F.}~\bibnamefont {Kuppers}}, \bibinfo {author} {\bibfnamefont
  {A.}~\bibnamefont {Soltani}},\ and\ \bibinfo {author} {\bibfnamefont {H.~G.}\
  \bibnamefont {Roskos}},\ }\bibfield  {title} {\bibinfo {title} {Direct
  near-field observation of surface plasmon polaritons on silver nanowires},\
  }\href@noop {} {\bibfield  {journal} {\bibinfo  {journal} {ACS Omega}\
  }\textbf {\bibinfo {volume} {4}},\ \bibinfo {pages} {21962} (\bibinfo {year}
  {2019})}\BibitemShut {NoStop}%
\bibitem [{Note1()}]{Note1}%
  \BibitemOpen
  \bibinfo {note} {Embedding MNWs in dielectric environments can extend the
  range to millimeters-known as long-range SPPs~\cite {berini2009long}.
  However, these structures are bulkier and reduce integration
  density.}\BibitemShut {Stop}%
\bibitem [{\citenamefont {Noginov}\ \emph {et~al.}(2009)\citenamefont
  {Noginov}, \citenamefont {Zhu}, \citenamefont {Belgrave}, \citenamefont
  {Bakker}, \citenamefont {Shalaev}, \citenamefont {Narimanov}, \citenamefont
  {Stout}, \citenamefont {Herz}, \citenamefont {Suteewong},\ and\ \citenamefont
  {Wiesner}}]{noginov2009demonstration}%
  \BibitemOpen
  \bibfield  {author} {\bibinfo {author} {\bibfnamefont {M.}~\bibnamefont
  {Noginov}}, \bibinfo {author} {\bibfnamefont {G.}~\bibnamefont {Zhu}},
  \bibinfo {author} {\bibfnamefont {A.}~\bibnamefont {Belgrave}}, \bibinfo
  {author} {\bibfnamefont {R.}~\bibnamefont {Bakker}}, \bibinfo {author}
  {\bibfnamefont {V.}~\bibnamefont {Shalaev}}, \bibinfo {author} {\bibfnamefont
  {E.}~\bibnamefont {Narimanov}}, \bibinfo {author} {\bibfnamefont
  {S.}~\bibnamefont {Stout}}, \bibinfo {author} {\bibfnamefont
  {E.}~\bibnamefont {Herz}}, \bibinfo {author} {\bibfnamefont {T.}~\bibnamefont
  {Suteewong}},\ and\ \bibinfo {author} {\bibfnamefont {U.}~\bibnamefont
  {Wiesner}},\ }\bibfield  {title} {\bibinfo {title} {Demonstration of a
  spaser-based nanolaser},\ }\href@noop {} {\bibfield  {journal} {\bibinfo
  {journal} {Nature}\ }\textbf {\bibinfo {volume} {460}},\ \bibinfo {pages}
  {1110} (\bibinfo {year} {2009})}\BibitemShut {NoStop}%
\bibitem [{\citenamefont {Huck}\ \emph {et~al.}(2009)\citenamefont {Huck},
  \citenamefont {Smolka}, \citenamefont {Lodahl}, \citenamefont {S{\o}rensen},
  \citenamefont {Boltasseva}, \citenamefont {Janousek},\ and\ \citenamefont
  {Andersen}}]{huck2009demonstration}%
  \BibitemOpen
  \bibfield  {author} {\bibinfo {author} {\bibfnamefont {A.}~\bibnamefont
  {Huck}}, \bibinfo {author} {\bibfnamefont {S.}~\bibnamefont {Smolka}},
  \bibinfo {author} {\bibfnamefont {P.}~\bibnamefont {Lodahl}}, \bibinfo
  {author} {\bibfnamefont {A.~S.}\ \bibnamefont {S{\o}rensen}}, \bibinfo
  {author} {\bibfnamefont {A.}~\bibnamefont {Boltasseva}}, \bibinfo {author}
  {\bibfnamefont {J.}~\bibnamefont {Janousek}},\ and\ \bibinfo {author}
  {\bibfnamefont {U.~L.}\ \bibnamefont {Andersen}},\ }\bibfield  {title}
  {\bibinfo {title} {Demonstration of quadrature-squeezed surface plasmons in a
  gold waveguide},\ }\href@noop {} {\bibfield  {journal} {\bibinfo  {journal}
  {Physical Review Letters}\ }\textbf {\bibinfo {volume} {102}},\ \bibinfo
  {pages} {246802} (\bibinfo {year} {2009})}\BibitemShut {NoStop}%
\bibitem [{\citenamefont {Di~Martino}\ \emph {et~al.}(2012)\citenamefont
  {Di~Martino}, \citenamefont {Sonnefraud}, \citenamefont {K{\'e}na-Cohen},
  \citenamefont {Tame}, \citenamefont {Ozdemir}, \citenamefont {Kim},\ and\
  \citenamefont {Maier}}]{di2012quantum}%
  \BibitemOpen
  \bibfield  {author} {\bibinfo {author} {\bibfnamefont {G.}~\bibnamefont
  {Di~Martino}}, \bibinfo {author} {\bibfnamefont {Y.}~\bibnamefont
  {Sonnefraud}}, \bibinfo {author} {\bibfnamefont {S.}~\bibnamefont
  {K{\'e}na-Cohen}}, \bibinfo {author} {\bibfnamefont {M.}~\bibnamefont
  {Tame}}, \bibinfo {author} {\bibfnamefont {S.~K.}\ \bibnamefont {Ozdemir}},
  \bibinfo {author} {\bibfnamefont {M.}~\bibnamefont {Kim}},\ and\ \bibinfo
  {author} {\bibfnamefont {S.~A.}\ \bibnamefont {Maier}},\ }\bibfield  {title}
  {\bibinfo {title} {Quantum statistics of surface plasmon polaritons in
  metallic stripe waveguides},\ }\href@noop {} {\bibfield  {journal} {\bibinfo
  {journal} {Nano Letters}\ }\textbf {\bibinfo {volume} {12}},\ \bibinfo
  {pages} {2504} (\bibinfo {year} {2012})}\BibitemShut {NoStop}%
\bibitem [{\citenamefont {Hong}\ \emph {et~al.}(2024)\citenamefont {Hong},
  \citenamefont {Dawkins}, \citenamefont {Bertoni}, \citenamefont {You},\ and\
  \citenamefont {Maga{\~n}a-Loaiza}}]{hong2024nonclassical}%
  \BibitemOpen
  \bibfield  {author} {\bibinfo {author} {\bibfnamefont {M.}~\bibnamefont
  {Hong}}, \bibinfo {author} {\bibfnamefont {R.~B.}\ \bibnamefont {Dawkins}},
  \bibinfo {author} {\bibfnamefont {B.}~\bibnamefont {Bertoni}}, \bibinfo
  {author} {\bibfnamefont {C.}~\bibnamefont {You}},\ and\ \bibinfo {author}
  {\bibfnamefont {O.~S.}\ \bibnamefont {Maga{\~n}a-Loaiza}},\ }\bibfield
  {title} {\bibinfo {title} {Nonclassical near-field dynamics of surface
  plasmons},\ }\href@noop {} {\bibfield  {journal} {\bibinfo  {journal} {Nature
  Physics}\ }\textbf {\bibinfo {volume} {20}},\ \bibinfo {pages} {830}
  (\bibinfo {year} {2024})}\BibitemShut {NoStop}%
\bibitem [{\citenamefont {Fasel}\ \emph {et~al.}(2005)\citenamefont {Fasel},
  \citenamefont {Robin}, \citenamefont {Moreno}, \citenamefont {Erni},
  \citenamefont {Gisin},\ and\ \citenamefont {Zbinden}}]{fasel2005energy}%
  \BibitemOpen
  \bibfield  {author} {\bibinfo {author} {\bibfnamefont {S.}~\bibnamefont
  {Fasel}}, \bibinfo {author} {\bibfnamefont {F.}~\bibnamefont {Robin}},
  \bibinfo {author} {\bibfnamefont {E.}~\bibnamefont {Moreno}}, \bibinfo
  {author} {\bibfnamefont {D.}~\bibnamefont {Erni}}, \bibinfo {author}
  {\bibfnamefont {N.}~\bibnamefont {Gisin}},\ and\ \bibinfo {author}
  {\bibfnamefont {H.}~\bibnamefont {Zbinden}},\ }\bibfield  {title} {\bibinfo
  {title} {Energy-time entanglement preservation in plasmon-assisted light
  transmission},\ }\href@noop {} {\bibfield  {journal} {\bibinfo  {journal}
  {Physical review letters}\ }\textbf {\bibinfo {volume} {94}},\ \bibinfo
  {pages} {110501} (\bibinfo {year} {2005})}\BibitemShut {NoStop}%
\bibitem [{\citenamefont {Fasel}\ \emph {et~al.}(2006)\citenamefont {Fasel},
  \citenamefont {Halder}, \citenamefont {Gisin},\ and\ \citenamefont
  {Zbinden}}]{fasel2006quantum}%
  \BibitemOpen
  \bibfield  {author} {\bibinfo {author} {\bibfnamefont {S.}~\bibnamefont
  {Fasel}}, \bibinfo {author} {\bibfnamefont {M.}~\bibnamefont {Halder}},
  \bibinfo {author} {\bibfnamefont {N.}~\bibnamefont {Gisin}},\ and\ \bibinfo
  {author} {\bibfnamefont {H.}~\bibnamefont {Zbinden}},\ }\bibfield  {title}
  {\bibinfo {title} {Quantum superposition and entanglement of mesoscopic
  plasmons},\ }\href@noop {} {\bibfield  {journal} {\bibinfo  {journal} {New
  Journal of Physics}\ }\textbf {\bibinfo {volume} {8}},\ \bibinfo {pages} {13}
  (\bibinfo {year} {2006})}\BibitemShut {NoStop}%
\bibitem [{\citenamefont {Berini}(2009)}]{berini2009long}%
  \BibitemOpen
  \bibfield  {author} {\bibinfo {author} {\bibfnamefont {P.}~\bibnamefont
  {Berini}},\ }\bibfield  {title} {\bibinfo {title} {Long-range surface plasmon
  polaritons},\ }\href@noop {} {\bibfield  {journal} {\bibinfo  {journal}
  {Advances in optics and photonics}\ }\textbf {\bibinfo {volume} {1}},\
  \bibinfo {pages} {484} (\bibinfo {year} {2009})}\BibitemShut {NoStop}%
\bibitem [{\citenamefont {Tasgin}(2026)}]{tasgin2026}%
  \BibitemOpen
  \bibfield  {author} {\bibinfo {author} {\bibfnamefont {M.~E.}\ \bibnamefont
  {Tasgin}},\ }\bibfield  {title} {\bibinfo {title} {Detecting nonclassicality
  in randomly-displaced copies of a squeezed state},\ }\href@noop {} {\bibfield
   {journal} {\bibinfo  {journal} {to appear in arXiv preprint}\ } (\bibinfo
  {year} {2026})},\ \bibinfo {note} {the Hamiltonian responsible for converting
  number-squeezing into quadrature-squeezing (and vice versa) is explicitly
  analyzed in this work. In addition, the constancy of $g^{(2)}(0)$ under
  purely lossy optical SPP propagation is derived in detail in the
  Appendix.}\BibitemShut {Stop}%
\bibitem [{PS_({\natexlab{a}})}]{PS_squeezing_detection}%
  \BibitemOpen
  \href@noop {} {}\bibinfo {note} {While squeezing detection via
  correlations~\cite{ou1987detection} has not yet been reported for plasmonic
  beam splitters, quantum interference effects—such as Hong–Ou–Mandel
  (HOM) interference—have been demonstrated in plasmonic
  systems~\cite{heeres2013quantum}.}\BibitemShut {Stop}%
\bibitem [{\citenamefont {G{\"u}ndogan}\ and\ \citenamefont
  {Tasgin}(2026)}]{gundogan2025enhanced}%
  \BibitemOpen
  \bibfield  {author} {\bibinfo {author} {\bibfnamefont {M.}~\bibnamefont
  {G{\"u}ndogan}}\ and\ \bibinfo {author} {\bibfnamefont {M.~E.}\ \bibnamefont
  {Tasgin}},\ }\bibfield  {title} {\bibinfo {title} {Quantum illumination with
  nonzero-mean signal-idler states via noise-enhanced heterodyne work
  extraction},\ }\href@noop {} {\bibfield  {journal} {\bibinfo  {journal}
  {arXiv preprint arXiv:2503.08248}\ } (\bibinfo {year} {2026})}\BibitemShut
  {NoStop}%
\bibitem [{\citenamefont {Huang}\ \emph {et~al.}(2023)\citenamefont {Huang},
  \citenamefont {Chi}, \citenamefont {Yang}, \citenamefont {Yang},
  \citenamefont {Zhai},\ and\ \citenamefont {Ou}}]{huang2023photonic}%
  \BibitemOpen
  \bibfield  {author} {\bibinfo {author} {\bibfnamefont {C.}~\bibnamefont
  {Huang}}, \bibinfo {author} {\bibfnamefont {H.}~\bibnamefont {Chi}}, \bibinfo
  {author} {\bibfnamefont {S.}~\bibnamefont {Yang}}, \bibinfo {author}
  {\bibfnamefont {B.}~\bibnamefont {Yang}}, \bibinfo {author} {\bibfnamefont
  {Y.}~\bibnamefont {Zhai}},\ and\ \bibinfo {author} {\bibfnamefont
  {J.}~\bibnamefont {Ou}},\ }\bibfield  {title} {\bibinfo {title} {{Photonic
  generation of dual-mode multi-format chirp microwave signals}},\ }\href@noop
  {} {\bibfield  {journal} {\bibinfo  {journal} {Applied Optics}\ }\textbf
  {\bibinfo {volume} {62}},\ \bibinfo {pages} {8224} (\bibinfo {year}
  {2023})}\BibitemShut {NoStop}%
\bibitem [{\citenamefont {Tan}\ \emph {et~al.}(2008)\citenamefont {Tan},
  \citenamefont {Erkmen}, \citenamefont {Giovannetti}, \citenamefont {Guha},
  \citenamefont {Lloyd}, \citenamefont {Maccone}, \citenamefont {Pirandola},\
  and\ \citenamefont {Shapiro}}]{tan2008quantum}%
  \BibitemOpen
  \bibfield  {author} {\bibinfo {author} {\bibfnamefont {S.-H.}\ \bibnamefont
  {Tan}}, \bibinfo {author} {\bibfnamefont {B.~I.}\ \bibnamefont {Erkmen}},
  \bibinfo {author} {\bibfnamefont {V.}~\bibnamefont {Giovannetti}}, \bibinfo
  {author} {\bibfnamefont {S.}~\bibnamefont {Guha}}, \bibinfo {author}
  {\bibfnamefont {S.}~\bibnamefont {Lloyd}}, \bibinfo {author} {\bibfnamefont
  {L.}~\bibnamefont {Maccone}}, \bibinfo {author} {\bibfnamefont
  {S.}~\bibnamefont {Pirandola}},\ and\ \bibinfo {author} {\bibfnamefont
  {J.~H.}\ \bibnamefont {Shapiro}},\ }\bibfield  {title} {\bibinfo {title}
  {Quantum illumination with gaussian states},\ }\href@noop {} {\bibfield
  {journal} {\bibinfo  {journal} {Physical Review Letters}\ }\textbf {\bibinfo
  {volume} {101}},\ \bibinfo {pages} {253601} (\bibinfo {year}
  {2008})}\BibitemShut {NoStop}%
\bibitem [{\citenamefont {Zhuang}(2021)}]{zhuang2021quantum}%
  \BibitemOpen
  \bibfield  {author} {\bibinfo {author} {\bibfnamefont {Q.}~\bibnamefont
  {Zhuang}},\ }\bibfield  {title} {\bibinfo {title} {Quantum ranging with
  gaussian entanglement},\ }\href@noop {} {\bibfield  {journal} {\bibinfo
  {journal} {Physical Review Letters}\ }\textbf {\bibinfo {volume} {126}},\
  \bibinfo {pages} {240501} (\bibinfo {year} {2021})}\BibitemShut {NoStop}%
\bibitem [{\citenamefont {Barzanjeh}\ \emph {et~al.}(2015)\citenamefont
  {Barzanjeh}, \citenamefont {Guha}, \citenamefont {Weedbrook}, \citenamefont
  {Vitali}, \citenamefont {Shapiro},\ and\ \citenamefont
  {Pirandola}}]{barzanjeh2015microwave}%
  \BibitemOpen
  \bibfield  {author} {\bibinfo {author} {\bibfnamefont {S.}~\bibnamefont
  {Barzanjeh}}, \bibinfo {author} {\bibfnamefont {S.}~\bibnamefont {Guha}},
  \bibinfo {author} {\bibfnamefont {C.}~\bibnamefont {Weedbrook}}, \bibinfo
  {author} {\bibfnamefont {D.}~\bibnamefont {Vitali}}, \bibinfo {author}
  {\bibfnamefont {J.~H.}\ \bibnamefont {Shapiro}},\ and\ \bibinfo {author}
  {\bibfnamefont {S.}~\bibnamefont {Pirandola}},\ }\bibfield  {title} {\bibinfo
  {title} {Microwave quantum illumination},\ }\href@noop {} {\bibfield
  {journal} {\bibinfo  {journal} {Physical Review Letters}\ }\textbf {\bibinfo
  {volume} {114}},\ \bibinfo {pages} {080503} (\bibinfo {year}
  {2015})}\BibitemShut {NoStop}%
\bibitem [{Note2()}]{Note2}%
  \BibitemOpen
  \bibinfo {note} {In principle in picosecond sequences, limited by the control
  solid-state device.}\BibitemShut {Stop}%
\bibitem [{\citenamefont {G{\"u}nay}\ \emph {et~al.}(2023)\citenamefont
  {G{\"u}nay}, \citenamefont {Das}, \citenamefont {Y{\"u}ce}, \citenamefont
  {Polat}, \citenamefont {Bek},\ and\ \citenamefont
  {Tasgin}}]{gunay2023demand}%
  \BibitemOpen
  \bibfield  {author} {\bibinfo {author} {\bibfnamefont {M.}~\bibnamefont
  {G{\"u}nay}}, \bibinfo {author} {\bibfnamefont {P.}~\bibnamefont {Das}},
  \bibinfo {author} {\bibfnamefont {E.}~\bibnamefont {Y{\"u}ce}}, \bibinfo
  {author} {\bibfnamefont {E.~O.}\ \bibnamefont {Polat}}, \bibinfo {author}
  {\bibfnamefont {A.}~\bibnamefont {Bek}},\ and\ \bibinfo {author}
  {\bibfnamefont {M.~E.}\ \bibnamefont {Tasgin}},\ }\bibfield  {title}
  {\bibinfo {title} {On-demand continuous-variable quantum entanglement source
  for integrated circuits},\ }\href@noop {} {\bibfield  {journal} {\bibinfo
  {journal} {Nanophotonics}\ }\textbf {\bibinfo {volume} {12}},\ \bibinfo
  {pages} {229} (\bibinfo {year} {2023})}\BibitemShut {NoStop}%
\bibitem [{\citenamefont {Tasgin}()}]{tasginSqueezingMultiplexed}%
  \BibitemOpen
  \bibfield  {author} {\bibinfo {author} {\bibfnamefont {M.~E.}\ \bibnamefont
  {Tasgin}},\ }\bibfield  {title} {\bibinfo {title} {Electrically controlled
  wavelength-multpilexed squeezing encoding},\ }\href@noop {} {\bibinfo
  {journal} {to appear in arXiv}\ }\BibitemShut {NoStop}%
\bibitem [{PS_({\natexlab{b}})}]{PS_g2_encoding}%
  \BibitemOpen
\bibfield  {journal} {  }\href@noop {} {}\bibinfo {note} {The $g^{(2)}(0)$
  encoding can also be implemented by applying a number-squeezing operation to
  an initially noisy state. The correlations generated by a single-mode state
  with $g^{(2)}(0)<1$ at the output of a plasmonic beam splitter can reveal the
  encoded data. Although $g^{(2)}(0)$ itself does not change during
  propagation, it becomes difficult to access when the remaining photon number
  is extremely small. If the preserved quantity ``were'' quadrature squeezing
  of a zero-mean state, it could still be recovered after long propagation, for
  example by applying the displacement operator $\hat{D}(\alpha)$ without
  altering the nonclassicality~\cite{SimonPRA1994}. In contrast, applying
  $\hat{D}(\alpha)$ to a state with $g^{(2)}(0)<1$ makes the nonclassical
  feature effectively invisible, as $g^{(2)}(0)$ approaches 1. Another option
  is to convert number squeezing into quadrature squeezing, or vice versa,
  using engineered interaction Hamiltonians without generating or removing
  nonclassicality~\cite{tasgin2026}. Although such Hamiltonians have mainly
  been studied in the high-photon-number regime~\cite{tasgin2026}, they may
  also be adapted for low-photon-number operation.}\BibitemShut {Stop}%
\bibitem [{\citenamefont {Ukhtary}\ and\ \citenamefont
  {Saito}(2020)}]{ukhtary2020surface}%
  \BibitemOpen
  \bibfield  {author} {\bibinfo {author} {\bibfnamefont {M.~S.}\ \bibnamefont
  {Ukhtary}}\ and\ \bibinfo {author} {\bibfnamefont {R.}~\bibnamefont
  {Saito}},\ }\bibfield  {title} {\bibinfo {title} {Surface plasmons in
  graphene and carbon nanotubes},\ }\href@noop {} {\bibfield  {journal}
  {\bibinfo  {journal} {Carbon}\ }\textbf {\bibinfo {volume} {167}},\ \bibinfo
  {pages} {455} (\bibinfo {year} {2020})}\BibitemShut {NoStop}%
\bibitem [{\citenamefont {Riedinger}\ \emph {et~al.}(2018)\citenamefont
  {Riedinger}, \citenamefont {Wallucks}, \citenamefont {Marinkovi{\'c}},
  \citenamefont {L{\"o}schnauer}, \citenamefont {Aspelmeyer}, \citenamefont
  {Hong},\ and\ \citenamefont {Gr{\"o}blacher}}]{riedinger2018remote}%
  \BibitemOpen
  \bibfield  {author} {\bibinfo {author} {\bibfnamefont {R.}~\bibnamefont
  {Riedinger}}, \bibinfo {author} {\bibfnamefont {A.}~\bibnamefont {Wallucks}},
  \bibinfo {author} {\bibfnamefont {I.}~\bibnamefont {Marinkovi{\'c}}},
  \bibinfo {author} {\bibfnamefont {C.}~\bibnamefont {L{\"o}schnauer}},
  \bibinfo {author} {\bibfnamefont {M.}~\bibnamefont {Aspelmeyer}}, \bibinfo
  {author} {\bibfnamefont {S.}~\bibnamefont {Hong}},\ and\ \bibinfo {author}
  {\bibfnamefont {S.}~\bibnamefont {Gr{\"o}blacher}},\ }\bibfield  {title}
  {\bibinfo {title} {Remote quantum entanglement between two micromechanical
  oscillators},\ }\href@noop {} {\bibfield  {journal} {\bibinfo  {journal}
  {Nature}\ }\textbf {\bibinfo {volume} {556}},\ \bibinfo {pages} {473}
  (\bibinfo {year} {2018})}\BibitemShut {NoStop}%
\bibitem [{\citenamefont {Abdi}\ \emph {et~al.}(2015)\citenamefont {Abdi},
  \citenamefont {Tombesi},\ and\ \citenamefont {Vitali}}]{abdi2015entangling}%
  \BibitemOpen
  \bibfield  {author} {\bibinfo {author} {\bibfnamefont {M.}~\bibnamefont
  {Abdi}}, \bibinfo {author} {\bibfnamefont {P.}~\bibnamefont {Tombesi}},\ and\
  \bibinfo {author} {\bibfnamefont {D.}~\bibnamefont {Vitali}},\ }\bibfield
  {title} {\bibinfo {title} {Entangling two distant non-interacting microwave
  modes},\ }\href@noop {} {\bibfield  {journal} {\bibinfo  {journal} {Annalen
  der Physik}\ }\textbf {\bibinfo {volume} {527}},\ \bibinfo {pages} {139}
  (\bibinfo {year} {2015})}\BibitemShut {NoStop}%
\bibitem [{\citenamefont {Balram}\ \emph {et~al.}(2017)\citenamefont {Balram},
  \citenamefont {Davan{\c{c}}o}, \citenamefont {Ilic}, \citenamefont {Kyhm},
  \citenamefont {Song},\ and\ \citenamefont {Srinivasan}}]{balram2017acousto}%
  \BibitemOpen
  \bibfield  {author} {\bibinfo {author} {\bibfnamefont {K.~C.}\ \bibnamefont
  {Balram}}, \bibinfo {author} {\bibfnamefont {M.~I.}\ \bibnamefont
  {Davan{\c{c}}o}}, \bibinfo {author} {\bibfnamefont {B.~R.}\ \bibnamefont
  {Ilic}}, \bibinfo {author} {\bibfnamefont {J.-H.}\ \bibnamefont {Kyhm}},
  \bibinfo {author} {\bibfnamefont {J.~D.}\ \bibnamefont {Song}},\ and\
  \bibinfo {author} {\bibfnamefont {K.}~\bibnamefont {Srinivasan}},\ }\bibfield
   {title} {\bibinfo {title} {Acousto-optic modulation and optoacoustic gating
  in piezo-optomechanical circuits},\ }\href@noop {} {\bibfield  {journal}
  {\bibinfo  {journal} {Physical Review Applied}\ }\textbf {\bibinfo {volume}
  {7}},\ \bibinfo {pages} {024008} (\bibinfo {year} {2017})}\BibitemShut
  {NoStop}%
\bibitem [{\citenamefont {Zhao}\ \emph {et~al.}(2023)\citenamefont {Zhao},
  \citenamefont {Bozkurt},\ and\ \citenamefont
  {Mirhosseini}}]{zhao2023electro}%
  \BibitemOpen
  \bibfield  {author} {\bibinfo {author} {\bibfnamefont {H.}~\bibnamefont
  {Zhao}}, \bibinfo {author} {\bibfnamefont {A.}~\bibnamefont {Bozkurt}},\ and\
  \bibinfo {author} {\bibfnamefont {M.}~\bibnamefont {Mirhosseini}},\
  }\bibfield  {title} {\bibinfo {title} {Electro-optic transduction in silicon
  via gigahertz-frequency nanomechanics},\ }\href@noop {} {\bibfield  {journal}
  {\bibinfo  {journal} {Optica}\ }\textbf {\bibinfo {volume} {10}},\ \bibinfo
  {pages} {790} (\bibinfo {year} {2023})}\BibitemShut {NoStop}%
\bibitem [{PS_({\natexlab{c}})}]{PS_room_T_operation}%
  \BibitemOpen
  \href@noop {} {}\bibinfo {note} {For instance, two optomechanically entangled
  phonons~\cite{riedinger2018remote,abdi2015entangling} can be converted into
  microwave photons via piezoelectric~\cite{balram2017acousto} or
  capacitive~\cite{zhao2023electro} couplings. Although direct observation of
  entanglement between the resulting microwave modes is obscured by thermal
  noise at such temperatures, as already demonstrated in QI experiments
  correlations (or extracted work~\cite{gundogan2025enhanced}) survives.
  Moreover, entanglement can still be inferred indirectly through the
  nonclassicality of the associated optical fields. Additionally,
  optical–microwave hybrid platforms allow the generation of correlated modes
  at room
  temperature~\cite{kitaeva2018generation,haase2019spontaneous,kuznetsov2020nonlinear}.}\BibitemShut
  {Stop}%
\bibitem [{PS_({\natexlab{d}})}]{PS_THz_advantange}%
  \BibitemOpen
  \href@noop {} {}\bibinfo {note} {Our technique is particularly advantageous
  for room-temperature devices operating in the THz regime. In this case, {\it
  the channel-induced thermal noise $\bar{n}_{\rm th}$ is much larger than the
  off-diagonal elements} of the covariance matrix $V_{\rm sqz}$ of a squeezed
  vacuum state. As a result, the squeezing signal is easily overwhelmed. In
  contrast, when squeezing is applied to a state that already contains
  room-temperature thermal noise~\cite{huang2023photonic}, the photon number
  satisfies $\bar{n}_{\rm p} = \bar{n}_{\rm th}$. In this case, the squeezing
  correlations are built on top of a noise level of the same order, which makes
  the encoded signal more robust against the channel noise.}\BibitemShut
  {Stop}%
\bibitem [{Note3()}]{Note3}%
  \BibitemOpen
  \bibinfo {note} {See Ref.~\cite {gundogan2025enhanced,huang2023photonic} for
  details of the technique studied in the context of quantum imaging and work
  extraction.}\BibitemShut {Stop}%
\bibitem [{PS_({\natexlab{e}})}]{PS_numner_of_photons}%
  \BibitemOpen
  \href@noop {} {}\bibinfo {note} {When the total photon numbers are compared
  directly, (i) squeezing a noisy state and (ii) sending a classical coherent
  signal may appear to require a similar number of photons. However, in the THz
  (or lower-frequency) regime, the photons involved in method (i) already exist
  naturally as thermal photons at room temperature~\cite{PS_room_T_operation}.
  In contrast, in method (ii) the photons forming the classical signal must be
  generated by a coherent source, which requires external energy input or work
  extraction. Even if the noisy initial state needs to be prepared—for
  example, by heating—the required energy is typically lower than the work
  needed to generate a coherent classical signal with the same photon number.
  Moreover, as discussed in the text, squeezing a noisy state is significantly
  more robust against channel noise than (iii) squeezing a vacuum state. The
  latter is also less practical at room temperature, since preparing a squeezed
  vacuum generally requires cryogenic
  cooling~\cite{chang2019quantum,assouly2023quantum} to suppress the thermal
  background.}\BibitemShut {Stop}%
\bibitem [{\citenamefont {Scully}\ and\ \citenamefont
  {Zubairy}(1997)}]{ScullyZubairyBook}%
  \BibitemOpen
  \bibfield  {author} {\bibinfo {author} {\bibfnamefont {M.~O.}\ \bibnamefont
  {Scully}}\ and\ \bibinfo {author} {\bibfnamefont {M.~S.}\ \bibnamefont
  {Zubairy}},\ }\href@noop {} {\bibinfo {title} {Quantum optics, {C}ambridge
  {U}niv. {P}ress}} (\bibinfo {year} {1997})\BibitemShut {NoStop}%
\bibitem [{\citenamefont {Gerry}\ and\ \citenamefont
  {Knight}(2023)}]{gerry2023introductory}%
  \BibitemOpen
  \bibfield  {author} {\bibinfo {author} {\bibfnamefont {C.~C.}\ \bibnamefont
  {Gerry}}\ and\ \bibinfo {author} {\bibfnamefont {P.~L.}\ \bibnamefont
  {Knight}},\ }\href@noop {} {\emph {\bibinfo {title} {Introductory quantum
  optics}}}\ (\bibinfo  {publisher} {Cambridge University Press},\ \bibinfo
  {year} {2023})\BibitemShut {NoStop}%
\bibitem [{\citenamefont {Guo}\ \emph {et~al.}(2023)\citenamefont {Guo},
  \citenamefont {Li}, \citenamefont {Wang}, \citenamefont {Meng}, \citenamefont
  {Zhao},\ and\ \citenamefont {Guo}}]{guo2023chaos}%
  \BibitemOpen
  \bibfield  {author} {\bibinfo {author} {\bibfnamefont {Y.}~\bibnamefont
  {Guo}}, \bibinfo {author} {\bibfnamefont {H.}~\bibnamefont {Li}}, \bibinfo
  {author} {\bibfnamefont {Y.}~\bibnamefont {Wang}}, \bibinfo {author}
  {\bibfnamefont {X.}~\bibnamefont {Meng}}, \bibinfo {author} {\bibfnamefont
  {T.}~\bibnamefont {Zhao}},\ and\ \bibinfo {author} {\bibfnamefont
  {X.}~\bibnamefont {Guo}},\ }\bibfield  {title} {\bibinfo {title} {Chaos with
  gaussian invariant distribution by quantum-noise random phase feedback},\
  }\href@noop {} {\bibfield  {journal} {\bibinfo  {journal} {Optics Express}\
  }\textbf {\bibinfo {volume} {31}},\ \bibinfo {pages} {31522} (\bibinfo {year}
  {2023})}\BibitemShut {NoStop}%
\bibitem [{\citenamefont {Chamoun}\ and\ \citenamefont
  {Digonnet}(2017)}]{chamoun2017aircraft}%
  \BibitemOpen
  \bibfield  {author} {\bibinfo {author} {\bibfnamefont {J.}~\bibnamefont
  {Chamoun}}\ and\ \bibinfo {author} {\bibfnamefont {M.~J.}\ \bibnamefont
  {Digonnet}},\ }\bibfield  {title} {\bibinfo {title}
  {Aircraft-navigation-grade laser-driven fog with gaussian-noise phase
  modulation},\ }\href@noop {} {\bibfield  {journal} {\bibinfo  {journal}
  {Optics Letters}\ }\textbf {\bibinfo {volume} {42}},\ \bibinfo {pages} {1600}
  (\bibinfo {year} {2017})}\BibitemShut {NoStop}%
\bibitem [{\citenamefont {Z{\~a}o}\ \emph {et~al.}(2009)\citenamefont
  {Z{\~a}o}, \citenamefont {Loss},\ and\ \citenamefont
  {Coelho}}]{zao2009design}%
  \BibitemOpen
  \bibfield  {author} {\bibinfo {author} {\bibfnamefont {L.}~\bibnamefont
  {Z{\~a}o}}, \bibinfo {author} {\bibfnamefont {G.}~\bibnamefont {Loss}},\ and\
  \bibinfo {author} {\bibfnamefont {R.}~\bibnamefont {Coelho}},\ }\bibfield
  {title} {\bibinfo {title} {Design and implementation of an optical gaussian
  noise generator},\ }\href@noop {} {\bibfield  {journal} {\bibinfo  {journal}
  {Optical Engineering}\ }\textbf {\bibinfo {volume} {48}},\ \bibinfo {pages}
  {085002} (\bibinfo {year} {2009})}\BibitemShut {NoStop}%
\bibitem [{\citenamefont {Chamoun}\ and\ \citenamefont
  {Digonnet}(2016)}]{chamoun2016pseudo}%
  \BibitemOpen
  \bibfield  {author} {\bibinfo {author} {\bibfnamefont {J.}~\bibnamefont
  {Chamoun}}\ and\ \bibinfo {author} {\bibfnamefont {M.~J.}\ \bibnamefont
  {Digonnet}},\ }\bibfield  {title} {\bibinfo {title}
  {Pseudo-random-bit-sequence phase modulation for reduced errors in a fiber
  optic gyroscope},\ }\href@noop {} {\bibfield  {journal} {\bibinfo  {journal}
  {Optics Letters}\ }\textbf {\bibinfo {volume} {41}},\ \bibinfo {pages} {5664}
  (\bibinfo {year} {2016})}\BibitemShut {NoStop}%
\bibitem [{\citenamefont {Capmany}\ and\ \citenamefont
  {Fern{\'a}ndez-Pousa}(2011)}]{capmany2011quantum}%
  \BibitemOpen
  \bibfield  {author} {\bibinfo {author} {\bibfnamefont {J.}~\bibnamefont
  {Capmany}}\ and\ \bibinfo {author} {\bibfnamefont {C.~R.}\ \bibnamefont
  {Fern{\'a}ndez-Pousa}},\ }\bibfield  {title} {\bibinfo {title} {Quantum
  modelling of electro-optic modulators},\ }\href@noop {} {\bibfield  {journal}
  {\bibinfo  {journal} {Laser \& Photonics Reviews}\ }\textbf {\bibinfo
  {volume} {5}},\ \bibinfo {pages} {750} (\bibinfo {year} {2011})}\BibitemShut
  {NoStop}%
\bibitem [{\citenamefont {Guha}\ and\ \citenamefont
  {Erkmen}(2009)}]{guha2009gaussian}%
  \BibitemOpen
  \bibfield  {author} {\bibinfo {author} {\bibfnamefont {S.}~\bibnamefont
  {Guha}}\ and\ \bibinfo {author} {\bibfnamefont {B.~I.}\ \bibnamefont
  {Erkmen}},\ }\bibfield  {title} {\bibinfo {title} {Gaussian-state
  quantum-illumination receivers for target detection},\ }\href@noop {}
  {\bibfield  {journal} {\bibinfo  {journal} {Physical Review A—Atomic,
  Molecular, and Optical Physics}\ }\textbf {\bibinfo {volume} {80}},\ \bibinfo
  {pages} {052310} (\bibinfo {year} {2009})}\BibitemShut {NoStop}%
\bibitem [{\citenamefont {Zhuang}\ \emph {et~al.}(2017)\citenamefont {Zhuang},
  \citenamefont {Zhang},\ and\ \citenamefont {Shapiro}}]{zhuang2017optimum}%
  \BibitemOpen
  \bibfield  {author} {\bibinfo {author} {\bibfnamefont {Q.}~\bibnamefont
  {Zhuang}}, \bibinfo {author} {\bibfnamefont {Z.}~\bibnamefont {Zhang}},\ and\
  \bibinfo {author} {\bibfnamefont {J.~H.}\ \bibnamefont {Shapiro}},\
  }\bibfield  {title} {\bibinfo {title} {Optimum mixed-state discrimination for
  noisy entanglement-enhanced sensing},\ }\href@noop {} {\bibfield  {journal}
  {\bibinfo  {journal} {Physical review letters}\ }\textbf {\bibinfo {volume}
  {118}},\ \bibinfo {pages} {040801} (\bibinfo {year} {2017})}\BibitemShut
  {NoStop}%
\bibitem [{\citenamefont {Shi}\ \emph {et~al.}(2023)\citenamefont {Shi},
  \citenamefont {Zhang},\ and\ \citenamefont {Zhuang}}]{shi2023fulfilling}%
  \BibitemOpen
  \bibfield  {author} {\bibinfo {author} {\bibfnamefont {H.}~\bibnamefont
  {Shi}}, \bibinfo {author} {\bibfnamefont {B.}~\bibnamefont {Zhang}},\ and\
  \bibinfo {author} {\bibfnamefont {Q.}~\bibnamefont {Zhuang}},\ }\bibfield
  {title} {\bibinfo {title} {Fulfilling entanglement’s optimal advantage via
  converting correlation to coherence},\ }in\ \href@noop {} {\emph {\bibinfo
  {booktitle} {CLEO: Fundamental Science}}}\ (\bibinfo {organization} {Optica
  Publishing Group},\ \bibinfo {year} {2023})\ pp.\ \bibinfo {pages}
  {FTh4A--6}\BibitemShut {NoStop}%
\bibitem [{\citenamefont {Reichert}\ \emph {et~al.}(2023)\citenamefont
  {Reichert}, \citenamefont {Zhuang}, \citenamefont {Shapiro},\ and\
  \citenamefont {Di~Candia}}]{reichert2023quantum}%
  \BibitemOpen
  \bibfield  {author} {\bibinfo {author} {\bibfnamefont {M.}~\bibnamefont
  {Reichert}}, \bibinfo {author} {\bibfnamefont {Q.}~\bibnamefont {Zhuang}},
  \bibinfo {author} {\bibfnamefont {J.~H.}\ \bibnamefont {Shapiro}},\ and\
  \bibinfo {author} {\bibfnamefont {R.}~\bibnamefont {Di~Candia}},\ }\bibfield
  {title} {\bibinfo {title} {Quantum illumination with a hetero-homodyne
  receiver and sequential detection},\ }\href@noop {} {\bibfield  {journal}
  {\bibinfo  {journal} {Physical Review Applied}\ }\textbf {\bibinfo {volume}
  {20}},\ \bibinfo {pages} {014030} (\bibinfo {year} {2023})}\BibitemShut
  {NoStop}%
\bibitem [{\citenamefont {Jeon}\ \emph {et~al.}(2025)\citenamefont {Jeon},
  \citenamefont {Kim}, \citenamefont {Kim}, \citenamefont {Kim}, \citenamefont
  {Jeong},\ and\ \citenamefont {Lee}}]{jeon2025single}%
  \BibitemOpen
  \bibfield  {author} {\bibinfo {author} {\bibfnamefont {S.}~\bibnamefont
  {Jeon}}, \bibinfo {author} {\bibfnamefont {J.}~\bibnamefont {Kim}}, \bibinfo
  {author} {\bibfnamefont {D.~Y.}\ \bibnamefont {Kim}}, \bibinfo {author}
  {\bibfnamefont {Z.}~\bibnamefont {Kim}}, \bibinfo {author} {\bibfnamefont
  {T.}~\bibnamefont {Jeong}},\ and\ \bibinfo {author} {\bibfnamefont {S.-Y.}\
  \bibnamefont {Lee}},\ }\bibfield  {title} {\bibinfo {title} {Single-mode
  phase-conjugate receiver for microwave quantum illumination with a lossy
  optical memory},\ }\href@noop {} {\bibfield  {journal} {\bibinfo  {journal}
  {Advanced Quantum Technologies}\ ,\ \bibinfo {pages} {2400627}} (\bibinfo
  {year} {2025})}\BibitemShut {NoStop}%
\bibitem [{\citenamefont {Karsa}\ \emph {et~al.}(2024)\citenamefont {Karsa},
  \citenamefont {Fletcher}, \citenamefont {Spedalieri},\ and\ \citenamefont
  {Pirandola}}]{karsa2024quantum}%
  \BibitemOpen
  \bibfield  {author} {\bibinfo {author} {\bibfnamefont {A.}~\bibnamefont
  {Karsa}}, \bibinfo {author} {\bibfnamefont {A.}~\bibnamefont {Fletcher}},
  \bibinfo {author} {\bibfnamefont {G.}~\bibnamefont {Spedalieri}},\ and\
  \bibinfo {author} {\bibfnamefont {S.}~\bibnamefont {Pirandola}},\ }\bibfield
  {title} {\bibinfo {title} {Quantum illumination and quantum radar: A brief
  overview},\ }\href@noop {} {\bibfield  {journal} {\bibinfo  {journal}
  {Reports on progress in physics}\ }\textbf {\bibinfo {volume} {87}},\
  \bibinfo {pages} {094001} (\bibinfo {year} {2024})}\BibitemShut {NoStop}%
\bibitem [{\citenamefont {Qian}\ \emph {et~al.}(2023)\citenamefont {Qian},
  \citenamefont {Xu}, \citenamefont {Zhu}, \citenamefont {Xu}, \citenamefont
  {Gao}, \citenamefont {Yakovlev}, \citenamefont {Liu}, \citenamefont {Zhu},\
  and\ \citenamefont {Wang}}]{qian2023quantum}%
  \BibitemOpen
  \bibfield  {author} {\bibinfo {author} {\bibfnamefont {G.}~\bibnamefont
  {Qian}}, \bibinfo {author} {\bibfnamefont {X.}~\bibnamefont {Xu}}, \bibinfo
  {author} {\bibfnamefont {S.-A.}\ \bibnamefont {Zhu}}, \bibinfo {author}
  {\bibfnamefont {C.}~\bibnamefont {Xu}}, \bibinfo {author} {\bibfnamefont
  {F.}~\bibnamefont {Gao}}, \bibinfo {author} {\bibfnamefont {V.}~\bibnamefont
  {Yakovlev}}, \bibinfo {author} {\bibfnamefont {X.}~\bibnamefont {Liu}},
  \bibinfo {author} {\bibfnamefont {S.-Y.}\ \bibnamefont {Zhu}},\ and\ \bibinfo
  {author} {\bibfnamefont {D.-W.}\ \bibnamefont {Wang}},\ }\bibfield  {title}
  {\bibinfo {title} {Quantum induced coherence light detection and ranging},\
  }\href@noop {} {\bibfield  {journal} {\bibinfo  {journal} {Physical Review
  Letters}\ }\textbf {\bibinfo {volume} {131}},\ \bibinfo {pages} {033603}
  (\bibinfo {year} {2023})}\BibitemShut {NoStop}%
\bibitem [{Note4()}]{Note4}%
  \BibitemOpen
  \bibinfo {note} {Once spaser technology reaches sufficient maturity, direct
  generation of nonclassical SPPs may also become possible~\cite
  {noginov2009demonstration,gunay2023demand}.}\BibitemShut {Stop}%
\bibitem [{\citenamefont {Han}\ \emph {et~al.}(2010)\citenamefont {Han},
  \citenamefont {Elezzabi},\ and\ \citenamefont {Van}}]{han2010wideband}%
  \BibitemOpen
  \bibfield  {author} {\bibinfo {author} {\bibfnamefont {Z.}~\bibnamefont
  {Han}}, \bibinfo {author} {\bibfnamefont {A.}~\bibnamefont {Elezzabi}},\ and\
  \bibinfo {author} {\bibfnamefont {V.}~\bibnamefont {Van}},\ }\bibfield
  {title} {\bibinfo {title} {Wideband y-splitter and aperture-assisted coupler
  based on sub-diffraction confined plasmonic slot waveguides},\ }\href@noop {}
  {\bibfield  {journal} {\bibinfo  {journal} {Applied Physics Letters}\
  }\textbf {\bibinfo {volume} {96}} (\bibinfo {year} {2010})}\BibitemShut
  {NoStop}%
\bibitem [{\citenamefont {Heeres}\ \emph {et~al.}(2013)\citenamefont {Heeres},
  \citenamefont {Kouwenhoven},\ and\ \citenamefont
  {Zwiller}}]{heeres2013quantum}%
  \BibitemOpen
  \bibfield  {author} {\bibinfo {author} {\bibfnamefont {R.~W.}\ \bibnamefont
  {Heeres}}, \bibinfo {author} {\bibfnamefont {L.~P.}\ \bibnamefont
  {Kouwenhoven}},\ and\ \bibinfo {author} {\bibfnamefont {V.}~\bibnamefont
  {Zwiller}},\ }\bibfield  {title} {\bibinfo {title} {Quantum interference in
  plasmonic circuits},\ }\href@noop {} {\bibfield  {journal} {\bibinfo
  {journal} {Nature nanotechnology}\ }\textbf {\bibinfo {volume} {8}},\
  \bibinfo {pages} {719} (\bibinfo {year} {2013})}\BibitemShut {NoStop}%
\bibitem [{\citenamefont {Kim}\ \emph {et~al.}(2002)\citenamefont {Kim},
  \citenamefont {Son}, \citenamefont {Bu{\v{z}}ek},\ and\ \citenamefont
  {Knight}}]{kim2002entanglement}%
  \BibitemOpen
  \bibfield  {author} {\bibinfo {author} {\bibfnamefont {M.}~\bibnamefont
  {Kim}}, \bibinfo {author} {\bibfnamefont {W.}~\bibnamefont {Son}}, \bibinfo
  {author} {\bibfnamefont {V.}~\bibnamefont {Bu{\v{z}}ek}},\ and\ \bibinfo
  {author} {\bibfnamefont {P.}~\bibnamefont {Knight}},\ }\bibfield  {title}
  {\bibinfo {title} {Entanglement by a beam splitter: Nonclassicality as a
  prerequisite for entanglement},\ }\href@noop {} {\bibfield  {journal}
  {\bibinfo  {journal} {Physical Review A}\ }\textbf {\bibinfo {volume} {65}},\
  \bibinfo {pages} {032323} (\bibinfo {year} {2002})}\BibitemShut {NoStop}%
\bibitem [{\citenamefont {Torrom{\'e}}\ and\ \citenamefont
  {Barzanjeh}(2024)}]{torrome2024advances}%
  \BibitemOpen
  \bibfield  {author} {\bibinfo {author} {\bibfnamefont {R.~G.}\ \bibnamefont
  {Torrom{\'e}}}\ and\ \bibinfo {author} {\bibfnamefont {S.}~\bibnamefont
  {Barzanjeh}},\ }\bibfield  {title} {\bibinfo {title} {Advances in quantum
  radar and quantum lidar},\ }\href@noop {} {\bibfield  {journal} {\bibinfo
  {journal} {Progress in Quantum Electronics}\ }\textbf {\bibinfo {volume}
  {93}},\ \bibinfo {pages} {100497} (\bibinfo {year} {2024})}\BibitemShut
  {NoStop}%
\bibitem [{\citenamefont {Ge}\ \emph {et~al.}(2015)\citenamefont {Ge},
  \citenamefont {Tasgin},\ and\ \citenamefont {Zubairy}}]{ge2015conservation}%
  \BibitemOpen
  \bibfield  {author} {\bibinfo {author} {\bibfnamefont {W.}~\bibnamefont
  {Ge}}, \bibinfo {author} {\bibfnamefont {M.~E.}\ \bibnamefont {Tasgin}},\
  and\ \bibinfo {author} {\bibfnamefont {M.~S.}\ \bibnamefont {Zubairy}},\
  }\bibfield  {title} {\bibinfo {title} {Conservation relation of
  nonclassicality and entanglement for gaussian states in a beam splitter},\
  }\href@noop {} {\bibfield  {journal} {\bibinfo  {journal} {Physical Review
  A}\ }\textbf {\bibinfo {volume} {92}},\ \bibinfo {pages} {052328} (\bibinfo
  {year} {2015})}\BibitemShut {NoStop}%
\bibitem [{\citenamefont {Braunstein}\ and\ \citenamefont
  {Van~Loock}(2005)}]{braunstein2005quantum}%
  \BibitemOpen
  \bibfield  {author} {\bibinfo {author} {\bibfnamefont {S.~L.}\ \bibnamefont
  {Braunstein}}\ and\ \bibinfo {author} {\bibfnamefont {P.}~\bibnamefont
  {Van~Loock}},\ }\bibfield  {title} {\bibinfo {title} {Quantum information
  with continuous variables},\ }\href@noop {} {\bibfield  {journal} {\bibinfo
  {journal} {Reviews of modern physics}\ }\textbf {\bibinfo {volume} {77}},\
  \bibinfo {pages} {513} (\bibinfo {year} {2005})}\BibitemShut {NoStop}%
\bibitem [{\citenamefont {Ou}\ \emph {et~al.}(1987)\citenamefont {Ou},
  \citenamefont {Hong},\ and\ \citenamefont {Mandel}}]{ou1987detection}%
  \BibitemOpen
  \bibfield  {author} {\bibinfo {author} {\bibfnamefont {Z.}~\bibnamefont
  {Ou}}, \bibinfo {author} {\bibfnamefont {C.}~\bibnamefont {Hong}},\ and\
  \bibinfo {author} {\bibfnamefont {L.}~\bibnamefont {Mandel}},\ }\bibfield
  {title} {\bibinfo {title} {Detection of squeezed states by cross
  correlation},\ }\href@noop {} {\bibfield  {journal} {\bibinfo  {journal}
  {Physical Review A}\ }\textbf {\bibinfo {volume} {36}},\ \bibinfo {pages}
  {192} (\bibinfo {year} {1987})}\BibitemShut {NoStop}%
\bibitem [{\citenamefont {Asb{\'o}th}\ \emph {et~al.}(2005)\citenamefont
  {Asb{\'o}th}, \citenamefont {Calsamiglia},\ and\ \citenamefont
  {Ritsch}}]{asboth2005computable}%
  \BibitemOpen
  \bibfield  {author} {\bibinfo {author} {\bibfnamefont {J.~K.}\ \bibnamefont
  {Asb{\'o}th}}, \bibinfo {author} {\bibfnamefont {J.}~\bibnamefont
  {Calsamiglia}},\ and\ \bibinfo {author} {\bibfnamefont {H.}~\bibnamefont
  {Ritsch}},\ }\bibfield  {title} {\bibinfo {title} {Computable measure of
  nonclassicality for light},\ }\href@noop {} {\bibfield  {journal} {\bibinfo
  {journal} {Physical review letters}\ }\textbf {\bibinfo {volume} {94}},\
  \bibinfo {pages} {173602} (\bibinfo {year} {2005})}\BibitemShut {NoStop}%
\bibitem [{\citenamefont {Duan}\ \emph {et~al.}(2000)\citenamefont {Duan},
  \citenamefont {Giedke}, \citenamefont {Cirac},\ and\ \citenamefont
  {Zoller}}]{duan2000inseparability}%
  \BibitemOpen
  \bibfield  {author} {\bibinfo {author} {\bibfnamefont {L.-M.}\ \bibnamefont
  {Duan}}, \bibinfo {author} {\bibfnamefont {G.}~\bibnamefont {Giedke}},
  \bibinfo {author} {\bibfnamefont {J.~I.}\ \bibnamefont {Cirac}},\ and\
  \bibinfo {author} {\bibfnamefont {P.}~\bibnamefont {Zoller}},\ }\bibfield
  {title} {\bibinfo {title} {Inseparability criterion for continuous variable
  systems},\ }\href@noop {} {\bibfield  {journal} {\bibinfo  {journal}
  {Physical review letters}\ }\textbf {\bibinfo {volume} {84}},\ \bibinfo
  {pages} {2722} (\bibinfo {year} {2000})}\BibitemShut {NoStop}%
\bibitem [{\citenamefont {Tasgin}(2019)}]{tasgin2019anatomy}%
  \BibitemOpen
  \bibfield  {author} {\bibinfo {author} {\bibfnamefont {M.~E.}\ \bibnamefont
  {Tasgin}},\ }\bibfield  {title} {\bibinfo {title} {Anatomy of entanglement
  and nonclassicality criteria},\ }\href@noop {} {\bibfield  {journal}
  {\bibinfo  {journal} {arXiv preprint arXiv:1901.04045}\ } (\bibinfo {year}
  {2019})}\BibitemShut {NoStop}%
\bibitem [{PS_({\natexlab{f}})}]{PS_squeezingNp}%
  \BibitemOpen
  \href@noop {} {}\bibinfo {note} {We remark that by telling ``$r=0.576$ amount
  of squeezing on a SMSNS'', we mean that the squeezing operator
  $\exp{[r(\hat{a}^2-H.c.)]}$ is applied on a single-mode noisy state
  $\bar{n}_{\rm p}$}\BibitemShut {NoStop}%
\bibitem [{Note5()}]{Note5}%
  \BibitemOpen
  \bibinfo {note} {There are three conceptually distinct methods for extracting
  squeezing information after SPP propagation: (i) Directly measuring noise in
  the single-mode SPP state after propagation over $L = 10L_0$; (ii) Splitting
  the same single-mode state using a beam splitter and then measuring output
  correlations ---this approach significantly reduces the number of copies
  needed; (iii) Using a squeezed noisy state (rather than a squeezed vacuum)
  and performing correlation measurements after beam splitting. This last
  method reduces the number of required measurements by a factor of $10^{-5}$
  to $10^{-4}$ compared to method (ii)~\cite
  {PS_numner_of_photons}.}\BibitemShut {Stop}%
\bibitem [{\citenamefont {Barzanjeh}\ \emph {et~al.}(2020)\citenamefont
  {Barzanjeh}, \citenamefont {Pirandola}, \citenamefont {Vitali},\ and\
  \citenamefont {Fink}}]{barzanjeh2020microwave}%
  \BibitemOpen
  \bibfield  {author} {\bibinfo {author} {\bibfnamefont {S.}~\bibnamefont
  {Barzanjeh}}, \bibinfo {author} {\bibfnamefont {S.}~\bibnamefont
  {Pirandola}}, \bibinfo {author} {\bibfnamefont {D.}~\bibnamefont {Vitali}},\
  and\ \bibinfo {author} {\bibfnamefont {J.~M.}\ \bibnamefont {Fink}},\
  }\bibfield  {title} {\bibinfo {title} {Microwave quantum illumination using a
  digital receiver},\ }\href@noop {} {\bibfield  {journal} {\bibinfo  {journal}
  {Science advances}\ }\textbf {\bibinfo {volume} {6}},\ \bibinfo {pages}
  {eabb0451} (\bibinfo {year} {2020})}\BibitemShut {NoStop}%
\bibitem [{\citenamefont {Simon}\ \emph {et~al.}(1994)\citenamefont {Simon},
  \citenamefont {Mukunda},\ and\ \citenamefont {Dutta}}]{SimonPRA1994}%
  \BibitemOpen
  \bibfield  {author} {\bibinfo {author} {\bibfnamefont {R.}~\bibnamefont
  {Simon}}, \bibinfo {author} {\bibfnamefont {N.}~\bibnamefont {Mukunda}},\
  and\ \bibinfo {author} {\bibfnamefont {B.}~\bibnamefont {Dutta}},\ }\bibfield
   {title} {\bibinfo {title} {Quantum-noise matrix for multimode systems: U(n)
  invariance, squeezing, and normal forms},\ }\href
  {https://doi.org/10.1103/PhysRevA.49.1567} {\bibfield  {journal} {\bibinfo
  {journal} {Phys. Rev. A}\ }\textbf {\bibinfo {volume} {49}},\ \bibinfo
  {pages} {1567} (\bibinfo {year} {1994})}\BibitemShut {NoStop}%
\bibitem [{\citenamefont {Kitaeva}\ \emph {et~al.}(2018)\citenamefont
  {Kitaeva}, \citenamefont {Kornienko}, \citenamefont {Leontyev},\ and\
  \citenamefont {Shepelev}}]{kitaeva2018generation}%
  \BibitemOpen
  \bibfield  {author} {\bibinfo {author} {\bibfnamefont {G.~K.}\ \bibnamefont
  {Kitaeva}}, \bibinfo {author} {\bibfnamefont {V.}~\bibnamefont {Kornienko}},
  \bibinfo {author} {\bibfnamefont {A.}~\bibnamefont {Leontyev}},\ and\
  \bibinfo {author} {\bibfnamefont {A.}~\bibnamefont {Shepelev}},\ }\bibfield
  {title} {\bibinfo {title} {Generation of optical signal and terahertz idler
  photons by spontaneous parametric down-conversion},\ }\href@noop {}
  {\bibfield  {journal} {\bibinfo  {journal} {Physical Review A}\ }\textbf
  {\bibinfo {volume} {98}},\ \bibinfo {pages} {063844} (\bibinfo {year}
  {2018})}\BibitemShut {NoStop}%
\bibitem [{\citenamefont {Haase}\ \emph {et~al.}(2019)\citenamefont {Haase},
  \citenamefont {Kutas}, \citenamefont {Riexinger}, \citenamefont {Bickert},
  \citenamefont {Keil}, \citenamefont {Molter}, \citenamefont {Bortz},\ and\
  \citenamefont {Von~Freymann}}]{haase2019spontaneous}%
  \BibitemOpen
  \bibfield  {author} {\bibinfo {author} {\bibfnamefont {B.}~\bibnamefont
  {Haase}}, \bibinfo {author} {\bibfnamefont {M.}~\bibnamefont {Kutas}},
  \bibinfo {author} {\bibfnamefont {F.}~\bibnamefont {Riexinger}}, \bibinfo
  {author} {\bibfnamefont {P.}~\bibnamefont {Bickert}}, \bibinfo {author}
  {\bibfnamefont {A.}~\bibnamefont {Keil}}, \bibinfo {author} {\bibfnamefont
  {D.}~\bibnamefont {Molter}}, \bibinfo {author} {\bibfnamefont
  {M.}~\bibnamefont {Bortz}},\ and\ \bibinfo {author} {\bibfnamefont
  {G.}~\bibnamefont {Von~Freymann}},\ }\bibfield  {title} {\bibinfo {title}
  {Spontaneous parametric down-conversion of photons at 660 nm to the terahertz
  and sub-terahertz frequency range},\ }\href@noop {} {\bibfield  {journal}
  {\bibinfo  {journal} {Optics Express}\ }\textbf {\bibinfo {volume} {27}},\
  \bibinfo {pages} {7458} (\bibinfo {year} {2019})}\BibitemShut {NoStop}%
\bibitem [{\citenamefont {Kuznetsov}\ \emph {et~al.}(2020)\citenamefont
  {Kuznetsov}, \citenamefont {Malkova}, \citenamefont {Zakharov}, \citenamefont
  {Tikhonova},\ and\ \citenamefont {Kitaeva}}]{kuznetsov2020nonlinear}%
  \BibitemOpen
  \bibfield  {author} {\bibinfo {author} {\bibfnamefont {K.~A.}\ \bibnamefont
  {Kuznetsov}}, \bibinfo {author} {\bibfnamefont {E.~I.}\ \bibnamefont
  {Malkova}}, \bibinfo {author} {\bibfnamefont {R.~V.}\ \bibnamefont
  {Zakharov}}, \bibinfo {author} {\bibfnamefont {O.~V.}\ \bibnamefont
  {Tikhonova}},\ and\ \bibinfo {author} {\bibfnamefont {G.~K.}\ \bibnamefont
  {Kitaeva}},\ }\bibfield  {title} {\bibinfo {title} {Nonlinear interference in
  the strongly nondegenerate regime and schmidt mode analysis},\ }\href@noop {}
  {\bibfield  {journal} {\bibinfo  {journal} {Physical Review A}\ }\textbf
  {\bibinfo {volume} {101}},\ \bibinfo {pages} {053843} (\bibinfo {year}
  {2020})}\BibitemShut {NoStop}%
\bibitem [{\citenamefont {Chang}\ \emph {et~al.}(2019)\citenamefont {Chang},
  \citenamefont {Vadiraj}, \citenamefont {Bourassa}, \citenamefont {Balaji},\
  and\ \citenamefont {Wilson}}]{chang2019quantum}%
  \BibitemOpen
  \bibfield  {author} {\bibinfo {author} {\bibfnamefont {C.}~\bibnamefont
  {Chang}}, \bibinfo {author} {\bibfnamefont {A.}~\bibnamefont {Vadiraj}},
  \bibinfo {author} {\bibfnamefont {J.}~\bibnamefont {Bourassa}}, \bibinfo
  {author} {\bibfnamefont {B.}~\bibnamefont {Balaji}},\ and\ \bibinfo {author}
  {\bibfnamefont {C.}~\bibnamefont {Wilson}},\ }\bibfield  {title} {\bibinfo
  {title} {Quantum-enhanced noise radar},\ }\href@noop {} {\bibfield  {journal}
  {\bibinfo  {journal} {Applied Physics Letters}\ }\textbf {\bibinfo {volume}
  {114}} (\bibinfo {year} {2019})}\BibitemShut {NoStop}%
\bibitem [{\citenamefont {Assouly}\ \emph {et~al.}(2023)\citenamefont
  {Assouly}, \citenamefont {Dassonneville}, \citenamefont {Peronnin},
  \citenamefont {Bienfait},\ and\ \citenamefont {Huard}}]{assouly2023quantum}%
  \BibitemOpen
  \bibfield  {author} {\bibinfo {author} {\bibfnamefont {R.}~\bibnamefont
  {Assouly}}, \bibinfo {author} {\bibfnamefont {R.}~\bibnamefont
  {Dassonneville}}, \bibinfo {author} {\bibfnamefont {T.}~\bibnamefont
  {Peronnin}}, \bibinfo {author} {\bibfnamefont {A.}~\bibnamefont {Bienfait}},\
  and\ \bibinfo {author} {\bibfnamefont {B.}~\bibnamefont {Huard}},\ }\bibfield
   {title} {\bibinfo {title} {Quantum advantage in microwave quantum radar},\
  }\href@noop {} {\bibfield  {journal} {\bibinfo  {journal} {Nature Physics}\
  }\textbf {\bibinfo {volume} {19}},\ \bibinfo {pages} {1418} (\bibinfo {year}
  {2023})}\BibitemShut {NoStop}%
\end{thebibliography}%

\end{document}